\documentclass[11pt]{emulateapj}
\usepackage{natbib}
\usepackage{amsmath,amsfonts,amssymb}
\usepackage{graphicx}
\usepackage{pslatex}
\def\idm#1{{\mbox{\scriptsize #1}}}

\newcommand\Ym{\langle Y\rangle}

\def\deg{{\rm o}}
\def\idm#1{{\mbox{\scriptsize #1}}}

\def\astrobj#1{#1\ }

\def\url#1{\texttt{#1}}
\newcommand\pstar{{\astrobj{PSR~1257+12}}}

\shortauthors{Go\'zdziewski, Konacki \& Wolszczan}
\begin{document}
\title{Long term stability and dynamical environment 
of the PSR 1257+12 planetary system}
\author{Krzysztof Go\'zdziewski\altaffilmark{1}}
\affil{Toru\'n Centre for Astronomy, N.~Copernicus University,
Gagarina 11, 87-100 Toru\'n, Poland}
\author{Maciej Konacki\altaffilmark{2}}
\affil{Department of Geological and Planetary Sciences, California
Institute of Technology, MS 150-21, Pasadena, CA 91125, USA\\
Nicolaus Copernicus Astronomical Center, Polish Academy of Sciences,
Rabia\'nska 8, 87-100 Toru\'n, Poland}
\author{Alex Wolszczan\altaffilmark{3}}
\affil{Department of Astronomy and Astrophysics, Penn State University, 
University Park, PA 16802, USA\\
Toru\'n Centre for Astronomy, Nicolaus Copernicus University
ul. Gagarina 11, 87-100 Toru\' n, Poland}
\altaffiltext{1}{e-mail: k.gozdziewski@astri.uni.torun.pl}
\altaffiltext{2}{e-mail: maciej@gps.caltech.edu}
\altaffiltext{3}{e-mail: alex@astro.psu.edu}

\begin{abstract}
We study the long-term dynamics of the \pstar planetary system. Using the 
recently determined accurate initial condition by Konacki \&
Wolszczan~(2003)  who derived the orbital inclinations and the absolute
masses of the planets B and C, we investigate the system stability by long-term,
1~Gyr direct integrations. No secular changes of the semi-major axes,
eccentricities and inclinations  appear during such an interval. This stable
behavior is confirmed with the fast indicator MEGNO. The analysis of the
orbital stability in the neighborhood of the  nominal initial condition
reveals that the \pstar system is localized in a wide stable region of the
phase space but close to a few weak 2 and 3-body mean motion resonances.
The long term  stability is additionally confirmed by a negligible exchange 
of the Angular Momentum Deficit between the innermost planet A and the pair 
of the outer planets B and C. An important feature of the system that helps 
sustain the stability is the secular apsidal resonance (SAR) between the 
planets B and C with the center of libration about $180^{\deg}$. We also 
find useful limits on the elements of the innermost planet~A which are 
otherwise unconstrained by the observations. Specifically, we find that 
the line of nodes of the planet A cannot be separated by more than about 
$\pm 60^{\circ}$ from the nodes of the bigger companions B and C. This limits 
the relative inclination of the orbit of the planet A to the mean orbital 
plane of the planets B and~C to moderate values. We also perform a preliminary  
study of the short-term dynamics of massless particles in the system.  
We find that a relatively extended stable zone exists between the planets A and B. 
Beyond the planet C, the stable zone appears already at distances 0.5~AU 
from the parent star. For moderately low eccentricities, beyond 1~AU, 
the motion of massless particles does not suffer from strong instabilities 
and this zone is basically stable, independent on the inclinations of 
the orbits of the test particles to the mean orbital plane of the system. 
It is an encouraging result supporting the search for a putative dust disk or 
a Kuiper belt, especially with the SIRTF mission. 
\end{abstract}

\keywords{celestial mechanics, stellar dynamics---methods: numerical, N-body
simulations---planetary systems---stars: individual (PSR 1257+12)}

\section{Introduction}

Up to date, among about 130 extrasolar planetary systems, the one around
\pstar remains the only one which contains Earth-sized
planets \citep{Wolszczan1992,Wolszczan1994,Konacki2003}. It has been
discovered with  the pulsar timing technique that relies on extremely precise
measurements  of the times of arrival (TOA) of pulsar pulses. Such a
technique in principle  allows a detection of companions as small as
asteroids. In the case of the  PSR~B1257+12 it enabled the detection of three
planets A, B and C with  the orbital periods of 25, 66 and 98 days and the
masses in the Earth-size  regime (see Table~1). Luckily, the two larger
planets have the mean motions  close to the 3:2 commensurability which result
in observable deviations  from a simple Keplerian description of the motion
\citep{Rasio1992, Malhotra1992,Peale1993,Wolszczan1994,Konacki1999}.
\cite{Konacki2003}  applied the secular orbital theory of the \pstar system
from \cite{Konacki2000} to determine the masses and absolute
inclinations of  the orbits of the planets B and C. Recently, the same idea
of incorporating the  effects of mutual interactions between planets to
remove the Doppler radial  velocity (RV) signal degeneracy present in the
$N$-Keplerian orbital model   (i.e., the undeterminancy of the system
inclination and the relative  inclination of the orbits) has also been
applied to study the strongly  interacting, resonant system around
Gliese~876~\citep{Laughlin2001,Rivera2001}. However, the accuracy of the
timing observations by far exceeds the precision  of the RV measurements. In
consequence, the accuracy of the \pstar fit  obtained by \cite{Konacki2003}
and hence the initial condition is superior to any initial
condition derived from the fits to RV observations of solar-type stars with
planets. Thus, the system constitutes a particularly convenient and
interesting subject for the studies of its dynamics.

There have been few and limited attempts to determine the long-term
stability  of the \pstar system. Using the resonance overlap criterion and
direct integrations, \cite{Rasio1992}, \cite{Malhotra1992} and
\cite{Malhotra1993a}  have found that the system disrupts on  $10^5$~yr
timescale if the masses exceed about 2-3 masses of Jupiter. Also 
\cite{Malhotra1992} and \cite{Malhotra1993} have estimated that if the 
masses were about $20-40$~M$_{\earth}$  then the system would be locked in
the exact 3:2 resonances which would lead to  the TOA signal very different
from what is actually observed. In this paper, we  investigate the stability
of the \pstar system in the Gyr-scale. We also perform  a dynamical
comparison of the \pstar system to the inner Solar System (ISS) as the
dynamics of the ISS is mostly driven by interactions between the  telluric
planets and is basically decoupled from the dynamics of the outer  Solar
System \citep{Laskar1994,Laskar1997}. Finally, we carry out  a preliminary
analysis of the stability of the orbits of massless particles (e.g. dust
particles or Kuiper belt type objects) to determine the zones  where such
particles could survive and possibly be detected.

\section{Numerical setup}

The stability analysis performed in this paper has mostly numerical character
and is understood in terms of the maximal Lyapunov Characteristic Number
(LCN). Hence we treat the quasi-periodic orbital motions (LCN $\simeq 0$) as
stable  and the chaotic ones (LCN $>0$) as unstable. In order to resolve the 
character of the orbits efficiently, we use the so-called fast indicator
MEGNO \cite[Mean Exponential Growth factor of Nearby
Orbits,][]{Cincotta2000}.  The MEGNO is directly related to  the LCN through
the linear relation $\Ym = a t  + b$. For quasi-periodic  motions $a \simeq 0$
and $b \simeq 2$ while for chaotic solutions  $a \simeq (\lambda/2)$ and
$b\simeq 0$ where $\lambda$ is the LCN of the orbit.  However, there is no
general relation between the magnitude of the LCN and the  macroscopic
changes of the orbital elements~\cite[e.g.,][]{Murison1994,Michtchenko2001}.
Even if the LCN is large, the system can stay bounded for a very long time.
More insight into the relation of the degree of the irregularity of  motion
and the macroscopic changes of the orbits one can obtain with the FFT techniques 
\citep{Michtchenko2001} and its refined variant, the Modified Fourier
Transform (MFT) developed by \cite{Laskar1993}. The MFT makes  it possible to
resolve the fundamental frequencies in a planetary system and to determine
their diffusion rates \citep{Robutel2001}. In our tests, we use the MFT to
verify the MEGNO integrations for massless particles  (Section~4) and to
identify the mean motion resonances (MMRs). During the computations of MEGNO,
for every planet $k=$A,B,C, and also for massless  particles ($k=0$) the
complex functions
$
     f_k(t_i) = a_k \exp \mbox{i}\lambda_k(t_i),
$
are computed at discrete times $t_i$,  where $t_{i+1}-t_{i} \simeq
10$~days, over the time $2T$ between $13,000~P_{\idm{C}}$ and 
$64,000~P_{\idm{C}}$ (where $P_{\idm{C}}$ is the orbital period of the
outermost  planet C). Here, $a_k$ denotes the semi-major axis of the relevant
planet and  $\lambda_k$ is its mean longitude. In general, for a quasi-periodic 
solution of a planetary system,
the frequency corresponding to the largest amplitude, $a_k^0$, is one of the
fundamental frequencies of motion, called the proper mean motion, $n_k$
\citep{Robutel2001}. The 2-body MMRs can be identified through  the condition
$
       q \nu_i - p \nu_j \simeq 0, \ i\neq j=\mbox{A,B,C,0}, 
$
where $p,q>0$ are prime integers. The MFT code gives us the rates
$\nu_i/\nu_j \simeq p/q $ and then $p$ and $q$ are found by the continued
fraction algorithm.  Such resonances will be labeled as 
$\mbox{P}_1q:\mbox{P}_2p$ where $\mbox{P}_1$ and $\mbox{P}_2$ denote
the bodies involved in the resonance. In this work we use the code which 
merges the MFT and MEGNO  methods. It incorporates the MFT code which has 
been kindly provided by David Nesvorny\footnote{\url{www.boulder.swri.edu/~davidn/fmft/}}.  

If a planetary system is strongly interacting, MEGNO enables us to detect
chaotic behavior typically over only $10^3$--$10^4$ orbital periods of the 
outermost planet. This feature is vital for examining large sets of the 
initial conditions \cite[see e.g.,][]{Gozdziewski2001a,Gozdziewski2003d}. 
However, due to very small masses in the \pstar system, the mutual perturbations 
between the planets are very week. In order to resolve the character of such orbits, 
much longer integration times are required. We have determined that 
the general purpose integrators (like the  Bulirsh-Stoer-Gragg, BSG) 
used to compute MEGNO are not efficient enough. Therefore, we have
developed a symplectic scheme of computing MEGNO based on the idea 
of \cite{Mikkola1999} (see also \cite{Gozdziewski2003b} and \cite{Cincotta2003}).
Instead of directly solving the variational equations of the perturbed Kepler 
problem, one can differentiate the symplectic leap-frog scheme~\citep{Wisdom1991} 
and compute the variations (and MEGNO) using the obtained symplectic tangent map. 
To use such a technique, a system of canonical variables is required. In these
variables, the planetary problem can be divided into integrable Keplerian motions 
and a perturbation which is integrable in the absence of the Keplerian part 
(usually it depends only on the coordinates). Our MEGNO code works internally 
in the Poincar\'e coordinates \citep{Laskar1995}. As the integrator core, the 
second- (LR2) and third- (LR3) order schemes by \cite{Laskar2001} are used. 
These integrators are much faster than the BSG algorithm. We can efficiently
perform intensive computations of MEGNO ($\simeq 10^4$ points per stability map) 
for $\simeq 10^6 P_{\idm{C}}$, keeping the fractional error of the total energy 
and the angular momentum below $10^{-10}$ when the 4~d step (for the LR3 integrator) 
is used. The tangent map approach is particularly suitable for the pulsar 
system due to typically small eccentricities of the investigated configurations.
The method will be described in detail elsewhere (Breiter and Gozdziewski).

However, one should be aware that even with such an extended integration
time-scale, in general we can only detect the strongest chaotic behavior
originating mostly from the mean motion resonances between the planets. Investigations
of subtle chaos originating in the secular system would require much longer
integration times, of the order of thousands of secular periods. Because
these periods are in the range $10^3$--$10^5$~yr (see Sect.~3) or longer,
the total integration time should be counted in tens of Myr. In such a case, 
due to the computational limitations, direct approach applied in this paper 
is no longer practically possible. 

In the numerical experiments, we use the initial condition given in 
Table~\ref{tab:tab1}. It represents one of the two orbital configurations 
obtained by \cite{Konacki2003} transformed to the classical astrocentric
elements. In some of the experiments these elements have been transformed 
to the canonical heliocentric elements inferred from the Poincar\'e coordinates.
In the second orbital configuration, the inclinations are such that $i_{\idm{B,C}}
\rightarrow 180^{\deg}-i_{\idm{B,C}}$. We assume that $i_{\idm{A}}$ obeys
the  same rule in that solution.  It follows that both these solutions are
dynamically equivalent and this reflects the fact that using the  secular
theory as the model of TOA measurements we are still not able to determine
the absolute direction of  the angular momentum vector. The TOA residuals
exclude opposite directions of  the outermost orbits \citep{Konacki2003}
hence the orbits are either prograde or retrograde.

\section{Stability of the \pstar planetary system}

We computed the MEGNO signature of the \pstar system over a few Myr 
($5\cdot 10^6$~$P_{\idm{C}}$). This test was repeated for many 
different, randomly selected initial variational vectors. The results 
of a few of the runs are shown in Figs~\ref{fig:fig1}fe. Both the regular 
evolution of MEGNO, $Y(t)$, as well as the quick convergence of its mean
value, $\Ym(t)$, to 2, indicate that the system is very close to a
quasi-periodic motion. Let us note, that our preliminary computations of MEGNO
with the BSG integrator revealed a very slow divergence of MEGNO after $0.5 \cdot
10^6$~$P_{\idm{C}}$ from the theoretical value of 2. It can be explained by
an accumulation of the numerical errors because this effect is absent in much
longer runs employing the symplectic code. Even if such a weak instability
were to be understood in terms of the chaotic behavior, it would correspond
to an extremely long Lyapunov time, $T_L = 1/\lambda$, comparable to
$10^9$~yr (estimated through the fit of the relation $\Ym =
(\lambda/2) t + b$ over $t\simeq 3.5\cdot 10^6$~yr).

In Fig.~\ref{fig:fig1}, we present the time-evolution of the canonical
heliocentric elements. Clearly, the motion of the planets B and C is tightly
coupled. This is illustrated in panels b and c which show the variations of
the eccentricities and inclinations. Fig.~\ref{fig:fig1}d demonstrates the
presence of a deep, so-called secular apsidal resonance (SAR), described by
the argument $\theta=\varpi_{\idm{B}}-\varpi_{\idm{C}}$ (where
$\varpi=\omega+\Omega$ is the longitude of periastron; $\Omega$ and $\omega$
are the longitude of ascending node and the argument of periastron of the
planet, respectively), with a small semi-amplitude of the librations,
$\simeq 50^{\deg}$, and the apsides on average anti-aligned. In the discovery
paper, \cite{Wolszczan1992} noticed the anti-alignment of the lines of
apsides while \cite{Rasio1992} and \cite{Malhotra1992} performed first
theoretical explorations of the resonance in the framework of the 
Laplace-Lagrange theory. 

The SAR was recently found in many extrasolar systems discovered by the
RV measurements. It is widely believed that the SAR, as the orbital state of
multi-planetary systems involving Jupiter-size planets, is crucial for
maintaining their long term stability \cite[for an overview see,][]{Ji2003}.
However in certain cases, the SAR should be understood as a typical feature of a planetary 
system when the secular angle $\theta$ oscillates about $0^{\circ}$ or $180^{\circ}$. 
In particular, this concerns the pulsar system: an inspection
of Table~3 shows that the proper frequencies of the pericenter motion,
$g_{\idm{p}}$, are not related through any simple linear combination fulfilling
the usual resonance rules. \cite{Lee2003} and \cite{Michtchenko2004}
show that the orbital configurations of two planets are generically 
stable if the system is far from strong mean motion resonances 
and collision zones, i.e., when the assumptions of the averaging 
theorem are fulfilled. In such a case, the secular character of 
the system depends on the variations of the angle~$\theta$.
Three different modes of its motion are possible: circulation, oscillations
around $0^{\circ}$ or $180^{\circ}$ and oscillations about $0^{\circ}$ in
the regime of large eccentricities. The first two modes are known from the 
classical Laplace-Lagrange linear secular theory~\citep{Dermott1999} while 
the third one was discovered by the cited authors. The third mode corresponds 
to the true, non-linear resonance and the secular system may be unstable in its
neighborhood. It should be noted that for the two first modes
there is no zero-frequency separatrix between the circulation
and oscillation regimes. In this sense, such a SAR is not a true resonance.

Due to a very small mass of the planet A, the pulsar system may be
approximated by a 3-body model with equal planetary masses, small eccentricities 
and inclinations. In such a case, \cite{Pauwels1983} showed that, at least in the
linear approximation, the SAR regime covers almost the entire phase space 
of the secular planetary system, the occurrence of a SAR is almost a certainty
and the system is stable. This could explain the stability of the pulsar system. 
However, some of its specific dynamical features break the assumptions of the 
secular model. Namely, the proximity of the system to the first-order 3:2 MMR 
and other mean motion resonances.   
In order to estimate the influence of the near 3:2~MMR on the secular solution, we
integrated the system over 10~Myr sampling the orbital elements every 100~yr.
Subsequently, the MFT was applied to analyze the complex signals
$e_{\idm{p}}\exp(\mbox{i}\varpi_{\idm{p}})$ and $\sin(i_{\idm{p}}/2)
\exp(\mbox{i}\Omega_{\idm{p}})$ derived from the heliocentric 
canonical elements. In this way, we can obtain an estimate of the precessional 
frequencies of the periastrons, $g_\idm{p}$, and the nodal lines, $s_{\idm{p}}$. 
These frequencies, together with the proper mean motions, $n_{\idm{p}}$,
compose the set of fundamental frequencies of the pulsar system. Their values 
and the corresponding periods are given in Table~3. The periastron frequencies, 
measured in arcsec/yr, are $\simeq  197.742$, $\simeq 43.881$ and $\simeq 13.157$. 
The first one is substantially different from its analytical approximation 
(in terms of the Laplace-Lagrange theory). This will lead to a fast discrepancy 
between the analytical and numerical solutions. Clearly, the classical secular 
theory has to be modified in order to account for the effects of
the near 3:2~MMR. Such a theory has been already developed by
\cite{Malhotra1989} for the  satellites of Uranus. It can also be applied 
to the \pstar system however it is somewhat beyond the scope of this paper. 

Using the symplectic integrator WHM~\citep{Wisdom1991} from the SWIFT
package~\citep{Duncan1994} as well as the LR2 integrator, we also
performed a few long term, 1~Gyr, integrations of the \pstar system. In the
first case, the time step equal to 1~d resulted in the fractional error of
the integrals of the angular momentum and the total energy at the level of 
$\simeq 10^{-10}$. The LR2 algorithm with the 4-day integration step
resulted in even higher accuracy. During the 1~Gyr interval, no signs of
instability are observed.  We have not detected any secular changes of 
the semi-major axes, eccentricities  and inclinations. Their values stay 
within the limits shown in  Fig.~\ref{fig:fig1} and the SAR persists with 
an unchanged amplitude of the librations (Fig.~\ref{fig:fig2}a) over 1~Gyr. 
 
Lest us note that from the 1~Gyr integrations it follows that the orbital 
elements of of the innermost planet A vary in a regular way (Fig.~\ref{fig:fig1}). 
We can also observe a time evolution of the argument  $\theta_1 =
\varpi_{\idm{A}}-\varpi{\idm{B}}$ as a semi-regular sequence of rotations
alternating with irregular ''librations''  about $180^{\deg}$. This effect is
preserved over the entire period of 1~Gyr (see Fig.~\ref{fig:fig2}c,d) and
may indicate that the orbit crosses the separatrix of a resonance. 
This is illustrated in Figs~\ref{fig:fig2}c,d  which show the
eccentricity of the planet A (multiplied by 10,000), $e_{\idm{A}}$, plotted
together with the argument $\theta_1$. Clearly, there is a correlation
between the librations of $\theta_1$ and small values of $e_{\idm{A}}$. We
also show $e_{\idm{C}}(\theta)$ and $e_{\idm{A}}(\theta_1)$ collected over
the 1~Gyr integration (Fig.~\ref{fig:fig2}e,f). The  first case
represents a trajectory in the resonant island of the SAR whereas  there is
no clear sign of librations in the $e_{\idm{A}}(\theta_1)$ plot. In fact,
this effect is only geometrical in nature and can be explained by estimating
the secular frequencies of the system using the well known Lagrange-Laplace
theory. If the MMRs are absent  and the disturbing function is expanded up to
the first order in the masses and to the second order in the eccentricities and
inclinations, the equations of the secular motion are integrable. Their solution,
relative to the eccentricities  and the longitudes of periastron, is given in
terms of the so called eccentricity vectors, $(h,k) = (e\sin
\varpi,e\cos\varpi)$, by \citep{Dermott1999}
\begin{eqnarray*}
  h_j(t) &=& \sum_{i} e_{j,i} \sin( g_i t + \beta_i ) \\
  k_j(t) &=& \sum_{i} e_{j,i} \cos( g_i t + \beta_i ),
\end{eqnarray*}
for every planet $j=$A,B,C. The constant amplitudes $e_{j,i}$ and secular
frequencies $g_i$ are the eigenvectors and eigenvalues of a matrix with
coefficients given explicitly in terms of the masses and constant semi-major 
axes of the planets. The scaling factors for the eigenvectors $e_{i,j}$ and 
the phases $\beta_i$ are determined by the initial condition. Geometrically, 
the time evolution of every eccentricity vector can be described as a
superposition  of the eigenmodes corresponding to $g_i$
\citep{Malhotra1993a}. The parameters  obtained for the \pstar system are
given in Table~\ref{tab:tab2}. They are  consistent with the results of
\cite{Malhotra1993} and \cite{Rasio1992} who analyzed the \pstar system
involving the two larger companions. They found  that the secular evolution
of the eccentricity vectors of the two outer planets  is almost entirely
driven by the eigenstate corresponding to $g_1$. Since we have $e_{B,1}
\simeq -e_{C,1}$ and the other components of $e_{j,i}$ are much smaller  and
almost equal, it follows that  $\omega_{\idm{B}} \simeq \omega_{\idm{C}}
+180^{\circ}$, which corresponds to the SAR of the planets B and C. For the
planet A, the eccentricity vector  is a superposition of all the three
eigenmodes with the leading second and third  mode having comparable
amplitudes. It appears that using the secular approximation, we can explain
the semi-librations  of $\theta_1$. The minima of $e_A$ occur when the modes
corresponding to the frequencies $g_2$ and $g_3$ are in anti-phase. Because
the amplitudes $e_{A,2}$ and $e_{A,3}$ have the same sign and similar
magnitudes while $e_{A,1}$ has a much smaller magnitude, the following
condition has to be satisfied to grant a minimum of $e_A$  
\[
  (g_2 t + \beta_2) - ( g_3 t + \beta_3 ) \simeq 180^{\circ}.
\]
Since $\beta_2-\beta_3 \simeq -180^{\deg}$ (see Table~\ref{tab:tab2}), we
have $(g_2-g_3) t = 360^{\deg}$ and the period of anti-alignment  is $P_{23} =
360^{\circ} /(g_2-g_3) \simeq 43,000$~yr. Curiously, from the relation between 
$g_i$, $g_1-7(g_2-g_3) \simeq 0.001^{\circ}/\mbox{yr}$, it follows that
this period is almost commensurate with $P_1$ (where $P_1 = 360^{\circ}/g_1$). 

Near the moment of anti-alignment, the eccentricity of the
planet A, $e_A$, is driven mostly by the $g_1$-mode that is in anti-phase
with the eccentricity vector for $e_{B}$. 
This leads to  the quasi-librations seen
in Fig.~\ref{fig:fig2} as they repeat with the same period as the variations
of $e_A$, $P_{23}$.

\subsection{Comparison to the inner Solar System}

The results of the 1~Gyr integration, expressed
in the canonical heliocentric elements, enable us to analyze
the time-evolution of the so-called Angular Momentum Deficit
\cite[AMD,][]{Laskar1997}: 
\[
 C = \sum_{p=\idm{A,B,C}} \frac{m_p m_{\star}}{m_p+m_{\star}}
        \sqrt{\mu_p a_p}(1-\sqrt{1-e_p^2}\cos i_p),
\] 
where $\mu_p=G( m_p + m_{\star})$, $G$ is the gravitational constant,
$a_p,e_p,i_p$ is the semi-major axis, eccentricity and inclination of 
a planet relative to the
invariant plane and $m_{\star}$ is the mass of the central body. The AMD
indicates the deviation of planetary orbits from a stable circular and
coplanar motion for which $C=0$. This quantity is preserved by the averaged
equations of motion \citep{Laskar1997}  and its stability  provides the
stability of the secular system in the absence of short-period resonances.
The AMD can be understood as the amplitude of the irregularity present in the
averaged system. Large values of AMD lead to a chaotic motion and for a certain
critical value to a crossing of the orbits and the disruption of a planetary
system \citep{Laskar1997,Laskar2000,Michtchenko2001}.

The \pstar system is close to the 3:2 commensurability. Its critical argument
circulates but it does not mean that the time-averaged effects of the
near-resonance vanish \citep{Malhotra1989,Malhotra1993}. Hence, the 
applicability of the AMD signature in the studies of the stability in the
real, unaveraged system can be problematic.  Instead, the integral
obtained by averaging the quasi-resonant system should be applied
\citep{Michtchenko2001}. Nevertheless, we have decided to calculate the AMD
and to examine its behavior in order to compare the results with those
obtained for the Solar system 
\cite[which contains two planets close to 5:2~MMR, see e.g.,][]{Ito2002}. 
The time evolution of AMD in the pulsar system is shown in Fig.~\ref{fig:fig2}. 
The AMD stays well bounded and very regular. There is very little exchange
of the AMD between the innermost planet A and the pair of the bigger
companions B and C. It suggests that the motion of the planet A is decoupled
from the dynamics of the planets B and C in the long-term scale. This
situation is qualitatively  different from the inner Solar system (ISS). The
AMD of the ISS is not strictly preserved  due to the perturbations of
Jupiter and Saturn, nevertheless  it can be considered roughly constant
\citep{Laskar1997,Ito2002}. \cite{Ito2002} published the
results of 5~Gr integrations of the ISS which reveal rapid AMD variations
of Mercury. They are much larger than the changes of AMD for the Venus-Earth pair.  
The variations of the AMD of Venus, Earth and Mars  are also irregular and
substantial. In fact, the ISS is chaotic having the Lyapunov time of about
$5$~Myr \citep{Laskar1994}. \cite{Laskar1994} found that  this chaos
is physically significant as it can lead to the ejection of Mercury  from the
inner ISS during a few Gyr. However, the source of chaos in the ISS is still
not well understood~\citep{Lecar2001}. In this sense, the character of the
motion of the \pstar system is quite different. The evolution of its AMD is
very regular in spite of smaller distances, larger masses and thus stronger
mutual interactions between the planets. Unlike Mercury's, the AMD of the
planet A is negligible when compared to the AMD of the B-C pair as it
contributes only about 1/1000 of the total value. On the other hand, the
orbital coupling of the pairs B-C and Venus-Earth \citep{Ito2002} is a
similar feature of the inner Solar and \pstar systems although it has a
different nature. The long term integrations by \cite{Ito2002} revealed an
anti-correlation between the changes of the orbital energies of Venus and 
Earth and, simultaneously, a correlation between the changes of their orbital
angular momenta and the eccentricities. These effects can be explained through the 
influence of the Jovian planets. Hence their dynamical source is external for the
coupled planets while the coupling of the B-C pair in the \pstar system is 
provided by the anti-aligned SAR. Finally, the chaotic orbital evolution of 
the ISS may significantly depend on the weak coupling with the outer planets. 
An equivalent effect is obviously absent in the \pstar system.

\subsection{Dynamical environment of the \pstar planetary system}

In the next set of experiments, we look at the initial condition in a global manner
in order to find out whether the current state of the system is robust to
the changes of the formal initial condition (IC). However, thanks to a very precise
determination of the initial condition such changes, if considered consistent
with the TOA measurements, are very limited. For example, the formal
$1\sigma$ error of the semi-major axes inferred from the parameter $x^0$
\citep{Konacki2003} is at  the level of $10^{-6}$~AU! Nevertheless, the
localization of the IC in the phase space (e.g. its proximity to
unstable regions) is critical to verify its character.
The examination of one isolated IC does not provide a definitive answer to
the question of stability.

\subsubsection{Mean motion resonances} 

We computed one-dimensional scans of MEGNO, $\Ym$, by changing the semi-major
axis of one planet and keeping the other orbital parameters fixed at the
values given in Table~\ref{tab:tab1}. The results of this experiment are
illustrated in Fig.~\ref{fig:fig3}. The MEGNO scans were computed with the 
resolution of $\simeq 5\cdot10^{-6}$~AU and $10^{-6}$~AU for close-up scans. 
They reveal a large number of spikes, some of them very close to the nominal positions 
of the planets marked with the large filled dots. Most of these spikes can be 
identified as 2-body MMRs between the planets.
Currently, the IC is well separated from low-order MMRs.  But some unstable
high-order MMRs appear close to the nominal IC. The most relevant ones seem
to be C31:B21~MMR (i.e. between the planets B and C, see the bottom panel in
Fig.~\ref{fig:fig3}) and C53:B36~MMR (shown in the scan for the planet~B).  Also
the planet A is close to B29:A11 MMR. Yet, such resonances are extremely narrow.
Their widths can be roughly estimated as less than $5\cdot 10^{-6}$~AU what
follows from their shapes and the resolution of the  MEGNO scans. 
It is unlikely that they can affect the regular motion of the system at its
nominal position. However, altering the semi-major axis of the planets by
small shifts, $\simeq 5\cdot 10^{-5}$~AU, would push the system into these unstable
regimes. In order to demonstrate it, we calculated MEGNO for a configuration 
corresponding to C31:B21 MMR (see the right 
column of Fig.~\ref{fig:fig4}). This resonance is the inclination-type MMR 
\citep[according to the terminology by][]{Peale1976}. Clearly, its critical argument
$\sigma=31\lambda_{\idm{C}}- 21\lambda{\idm{B}}-10\Omega_{\idm{B}}$ exhibits
a sequence of alternating librations and circulation giving rise to the
unstable behavior. Let us note, that the chaos is formally extremely
strong. The Lyapunov time, estimated by the linear fit to the MEGNO, is only
about 3000~yr (see Fig.~\ref{fig:fig4}a).
 
The 2-body MMRs cannot explain all the peaks of MEGNO present in the scans. 
Some of them seem to be related to the 3-body resonances involving all three planets. 
The importance of 3-body MMRs was studied by \cite{Nesvorny1998,Morbidelli1999}, 
\cite{Murray1998} and \cite{Murray1999}. They found that weak 3-body MMRs can strongly 
influence asteroidal motion and explain the short Lyapunov times which cannot be 
understood if only 2-body MMRs of high-order are considered. \cite{Murray1999} proved 
that the chaotic behavior of the Outer Solar system is governed by the overlapping 3-body 
MMRs involving Jupiter, Saturn, and Uranus. It is obviously interesting to
verify if some of the instabilities visible in the $a$-scans can be related to such
3-body resonances.

A 3-body resonance can be defined by the following
condition~\citep{Nesvorny1998}:
\[
        i_{\idm{A}} {\dot \lambda_{\idm{A}}} + i_{\idm{B}} 
	{\dot \lambda_{\idm{B}}}
	+ i_{\idm{C}} {\dot \lambda_{\idm{C}}} \simeq 0,
\]
where ${\dot \lambda{\idm{p}}}$ is the mean motion of the given planet. A
critical argument of such resonance is a linear  combination of the
longitudes, arguments of pericenters and nodal longitudes:
\begin{eqnarray*}
\sigma_{i_{\idm{A}}:i_{\idm{B}}:i_{\idm{C}}} & = &
i_{\idm{A}} \lambda_{\idm{A}} + 
i_{\idm{B}} \lambda_{\idm{B}} + 
i_{\idm{C}} \lambda_{\idm{C}} + \\
&& p_{\idm{A}} \varpi_{\idm{A}}  + 
p_{\idm{B}} \varpi_{\idm{B}}  + 
p_{\idm{C}} \varpi_{\idm{C}}  +
q_{\idm{A}} \Omega_{\idm{A}}  + 
q_{\idm{B}} \Omega_{\idm{B}}  + 
q_{\idm{C}} \Omega_{\idm{C}} 
\end{eqnarray*}
whose integer coefficients $i_{\idm{p}},p_{\idm{p}},q_{\idm{p}}$ fulfill the 
d'Alambert rule	$\sum_{\idm{p=A,B,C}}
(i_{\idm{p}}+p_{\idm{p}}+q_{\idm{p}})=0$ and the usual requirement of
rotational symmetry. Due to the small ratios of the secular
frequencies to the mean motions in the \pstar system, $\simeq 10^{-5}$, the
resonance condition may be approximated by 
$
i_{\idm{A}}{n_{\idm{A}}} + 
i_{\idm{B}}{n_{\idm{B}}} + 
i_{\idm{C}}{n_{\idm{C}}} \simeq 0.
$
For example, one of the MEGNO peaks for $a_{\idm{C}} = 0.46601~AU$, can 
be explained as the combination of the proper mean motions
  $3 n_{\idm{A}} - 14 n_{\idm{B}} + 9 n_{\idm{C}} \simeq
0.01^{\circ}/\mbox{yr}$. Hence it corresponds to the 3-body MMR~3:-14:9. It
is separated from the the nominal position of the planet C by only $\simeq
3\cdot 10^{-5}$~AU which is at the $3\sigma$ error level of $a_{\idm{C}}$. 
The time evolution of the critical argument,
$\sigma_{3:-14:9}$, shown in the left column of Fig.~\ref{fig:fig4}, confirms
the slow changes of this angle as well as a presence of quasi-librations
alternating with circulations. This indicates a chaotic evolution.
The MEGNO computed for the corresponding initial condition does not grow
fast, at least over a few Myr used in this test. This mean that the
chaos is moderate. Nevertheless, the long term  stability of the system 
(likely, the chaos would affect mostly the motion of the planet A) could be
confirmed only by direct numerical integrations. Our 1~Gyr integrations 
for the ICs of the two MMRs near the planet C do not reveal any
secular changes of the elements. Likely, in order to properly investigate subtle 
effects of these resonances and their influence on the system, a
refined model of the motion should be used. Such model should incorporate 
the effects of general relativity, a possible error in the pulsar mass and 
other factors, like the dependence on the observationally unconstrained 
orbital elements of the planet~A. 

Other low-order 3-body resonances that can be identified in the
Fig.~\ref{fig:fig3} are 1:-2:1, 1:-4:2, -2:10:7 (almost overlapping with
B71:A27), 3:-16:12, 1:-6:5 (see the $a_{\idm{A}}$-scan). Finally, also
an overlapping of 2-body MMRs is possible. One such example is marked in the
$a_{\idm{B}}$-scan where in the neighborhood of the 3:2~MMR, a sharp peak of
MEGNO at the simultaneous position of B60:A23 and C76:B51 MMRs is present.
All these MMRs are very narrow since the width of 3-body MMRs is
proportional to the masses in the second order and these 2-body MMRs are 
of high order. They are relatively distant from the nominal positions 
of the planets thus it is unlikely that they can affect the motion of 
the system.  

Finally, Fig.~\ref{fig:fig4} shows the effect of varying the assumed mass of the
neutron  star on the determination of the semi-major axis of the outermost
planet and the localization of the resonances.  Obviously, the sequence of
MMRs does not change. A different mass of the pulsar can lead to a
substantial shift of  the initial  semi-major axis (in our case,
$a_{\idm{C}}$) and the weak unstable resonances, like the  C31:B21 MMR, may
end up much closer to the nominal positions of the planets (for smaller than
canonical mass of the parent star, $M_{\idm{psr}}=1.4M_{\sun}$).

\subsubsection{
Stability  in the $(a_{\idm{p}},e_{\idm{p}})$- and $(e_{\idm{B}},e_{\idm{C}})$-planes}

The results of the 2-dimensional MEGNO analysis are shown in Fig.~\ref{fig:fig6}. 
In these maps we analyze the influence of the initial eccentricities of the 
planets on their motion. We used the initial conditions from Table~1 
and changed those elements that correspond to the coordinates in the stability maps.
Due to a lower resolution (400 points in the semi-major axis range),  
some features present in the one-dimensional scans are not so clearly visible.
Nevertheless, in the $(a_{\idm{B}},e_{\idm{B}})$-plane (the middle panel), 
the thin strips of the MMRs can be easily identified.  These maps
additionally reveal the width of the strongest MMRs (like C3:B2, C13:B9
and C10:B7) and the border of the global instability. These stability maps confirm
that the nominal positions of the two bigger planets, marked with filled
circles, are far from strong instabilities of the motion. Also in the 
$(e_{\idm{B}},e_{\idm{C}})$-plane, the nominal IC lies in a wide
stable region, far from zones of chaotic motion (see Fig.~\ref{fig:fig7}). 
The $(a_{\idm{A}},e_{\idm{A}})$map for the planet A shows a dense net
of narrow unstable regions, some of which are close to the nominal position of
the planet. They have been already identified in the 1D scans for $a_{\idm{A}}$. 
Let us note that similar maps were computed by \cite{FerrazMello2002}
for the planet~B with their FFT fast indicator. Their results are generally
consistent with ours even though they used a different initial condition of 
the pulsar system.

During the MEGNO integrations we have also computed the maximum value of the
SAR argument $\theta = \varpi_{\idm{B}} - \varpi_{\idm{C}}$ with respect to
the libration  center of $180^{\deg}$. It enables us to estimate the
semi-amplitude of librations and  the extent of the SAR in the space of the
scanned orbital parameters. The result of such  an experiment conducted in
the $(a_{\idm{B}},e_{\idm{B}})$ plane are shown in  Fig.~\ref{fig:fig8}. It
uncovers an extended zone of the SAR with the smallest  semi-amplitude of
$\theta$ in the vicinity of the nominal initial condition of the \pstar 
system. The near 3:2 MMR is also clearly present in this map.

\subsubsection{Limits on the unconstrained parameters of planet A}

So far, after \cite{Konacki2003}, we were using the average inclination of 
the orbits of the planets B and C as the orbital inclination of the 
planet~A.  However, we can try to verify whether the dynamics can provide
any limits on the inclination and hence mass of the innermost planet A. To
this end, we computed MEGNO for the inclination of the planet A in the range
$[30^{\deg},70^{\deg}]$  and  the position of its nodal line in the range
$[-90^{\deg}:90^{\deg}]$, with the resolution of $1^{\circ}$ in both
coordinates. Thus we varied both the mass (to preserve 
$m{\idm{A}}\sin i_{\idm{A}}$ determined from the TOAs) 
as well as the relative
inclination of the planet A to the orbital planes of~B and~C. The result 
of this experiment is shown in Fig.~\ref{fig:fig9}. The MEGNO scan
(the left panel of Fig.~\ref{fig:fig9}) reveals a well defined stable region. 
The assumed position of the planet A is in the center of this zone. 
To obtain this picture, an extended integration time span of $10^6 P_{\idm{C}}$ 
was used. This choice was dictated by the analysis of a few orbits of A 
which are substantially inclined to the orbital planes of B and C. 
An example corresponding to the initial $\Omega{\idm{A}}=-75^{\circ}$ 
(thus located in the unstable region) is shown in Fig.~\ref{fig:fig10}. 
For the integration time up to about 0.5~Myr ($\simeq 5\cdot 10^5 P_{\idm{C}}$) 
MEGNO stays close to~2 but then it suddenly grows, indicating chaotic behavior.
To explain its source, we computed the eccentricity, inclination
$i^{\idm{rel}}_{\idm{A}}$ and the argument of pericenter, $g_{\idm{A}}$, of
the planet A relative to the invariant plane. Evolution of these elements is
shown in Fig.\ref{fig:fig10}. Clearly, both $e_{\idm{A}}$ and
$i^{\idm{rel}}_{\idm{A}}$  exhibit long-term, large amplitude oscillations
which are exactly in an anti-phase. The period of these oscillation is 
relatively very long, of the order of $10^4$~yr. Simultaneously,  
$g_{\idm{A}}$  temporarily librates about $90^{\circ}$ or $270^{\circ}$.  
These librations indicate that the pericenter precession of A~stops. 
Such features are typical for the well known Kozai resonance~\citep{Kozai1962} 
found for highly inclined asteroids in the Solar system. The argument of pericenter 
$g_{\idm{A}}$ can be treated as the critical argument of this resonance. Because
librations are followed by circulations of the critical argument, the orbit 
crosses the separatrix which explains why the motion eventually becomes chaotic. 

The effects of the Kozai resonance are illustrated  in the right panel of 
the $e^{\mbox{max}}_{\idm{A}}$-map (the maximal eccentricity attained by the 
planet A during the integration time) in Fig.~\ref{fig:fig9}. 
The sharply ending zone of moderate eccentricities is narrower than the
stability region. We should be aware that in the transient areas the
integration time could still be to short to detect the irregular motion.
Note that around the border of the stable zone, $e_{\idm{A}}$ can be as
large as 0.6--0.8. This resonance puts significant limits on the position 
of the nodal line of the planet A as well as its relative inclination 
to the orbital planes of B and C. Assuming that the motion of all three planets 
is in the same direction, the nodal line of A cannot be separated by 
more than about $\pm 60^{\circ}$ from the nodes of B and C and the relative 
inclination should not exceed $\simeq 45^{\circ}$. Otherwise, the eccentricity 
of A would be excited to large values, $>0.7$ over only $10^5$~yr and 
for $e_{\idm{A}} \simeq 0.8$ the orbits of A and B would cross or close 
encounters would become possible.

\section{Dynamics of massless particles}

While investigating the dynamics of extrasolar planetary systems one can ask
whether the minor bodies can survive in the gravitational environment of the
giant primary bodies. These can be small telluric planets in the systems
with Jupiter-like planets. We can ask the same question about
asteroids, cometary bodies or dust particles in the \pstar system. Because the 
motion of the planets appears to be strictly stable, we can consider a simplified,
restricted model. In this model one assumes that a probe mass moves in a
gravitational tug of the primaries but does not influence their motion. In
the simplest case such a model is well known as the restricted, planar
circular 3-body problem~\citep{Dermott1999}. This simplification helps us to explore the
phase space of the planetary system with reasonable computational effort. In
this model, the MEGNO indicator and the MFT are evaluated only for the probe mass.
In the numerical experiments, we investigate the orbital stability of
massless particles in a few regions of the orbital space: between the planets A
and B (region I), B and C (region II) and beyond the planet C (region III).   

In the first series of numerical tests, we vary the initial semi-major axis
of the  probe mass, $a_0$, and its eccentricity, $e_{\idm{0}}$, while the
initial inclination  is constant, $i_{\idm{0}}=50^{\deg}$. Thus the tested
orbits are  slightly inclined to both orbital planes of the planets B
($i_{\idm{B}}=53^{\deg}$)  and C ($i_{\idm{C}}=47^{\deg}$). The angular
variables of the massless particles are set to
$\Omega_{\idm{0}}=\varpi_{\idm{0}}=M_{\idm{0}}=0^{\deg}$.   This way we
essentially follow the remarkable work of \cite{Robutel2001} who 
investigated the short-term dynamics of massless particles in the Solar
System using  the MFT. These authors argue that the global picture of the
motion in the restricted problem does not qualitatively depend on the initial
phases of the test  particle. The initial plane $(a_0,e_0)$ is representative
for the dynamics because  it crosses all resonances. By changing the initial
orbital phases of the massless particles, the width of the  resonances may
vary but their influence on the motion can still be detected.

In order to check the MEGNO signatures (e.g., to verify whether the
integration time  is not too short), during the integrations, we compute the
diffusion rate,  $\sigma_0 = 1 - n^{(2)}/n^{(1)}$, where $n^{(1)}$ and
$n^{(2)}$ are the  mean motion  frequencies obtained over the
intervals $[0,T]$ and $[T,2T]$, respectively,  where $T$ (which we will  call
the base period from hereafter) has been set accordingly to the investigated
range  of the semi-major axes. The close encounters with the planets are
controlled. We  assume that if the distance between the probe mass and a
planet is less than the Hill  radius, $r_H = [m_p/(3 m_{\star})]^{(1/3)}
a_p$, then the particle has collided with the  planet or its motion become
strongly chaotic. 

Within this numerical setup, in the region II we have not detected any stable
motions.  The region~I turns out to be a much more interesting one. The
results are shown in the MEGNO  and $\sigma$ maps in Fig.~\ref{fig:fig11}
(left panels for the region I, middle and right panels for the region III).
The integrations have been carried out for the  total time, $2T$, of about
32,000~$P_{\idm{C}}$. In both maps we can clearly recognize  the
planetary-crossing lines marked by large values of both indicators which
correspond to  strongly chaotic motions. These lines (white lines in
Fig.~\ref{fig:fig11}) are the solutions  to the equations
$a_p(1+e_p)=a_0(1-e_0)$ for $a_0>a_p$ and $a_p(1-e_p)=a_0(1+e_0)$ for 
$a_0<a_p$ where $a_p,e_p,a_0,e_0$ denote the initial semi-major axis and the
eccentricity  of a planet and a test particle, respectively.

In this area and above the collision lines, the particles are scattered 
mostly by close-encounters and collisions with the planets A and B.
Remarkably, a stable area exists under the  crossing-lines for the planets A
and B. It is divided by a number of MMRs with the  planets. In order to
verify this result, we have integrated the motion of 200 particles  spread
over $[0.19,0.35]$~AU with the same resolution in the semi-major axis,
$a_0$,  as in the 2D MEGNO test. These integrations were continued up to
15~Myr (here, we used the  RMVS3 integrator from the SWIFT package). A
resulting one dimensional scan of MEGNO over  $a_{\idm{0}}$ is shown in
Fig.~\ref{fig:fig12}. In the regions classified with MEGNO as  regular, the
massless particles have mostly survived during the integration time while in 
the chaotic or close to chaotic areas, they have been quickly removed or
collided with the  planets (in this test the same criterion of 1~Hill radii
for a collision  event is used and in such a case we set $\log \sigma_0 =
1$). Some discrepancies  are most likely related to the different integrators
used in the experiment; the Bulirsh-Stoer integrator follows the orbits of
massless particles much more accurately that the symplectic  integrator.
This experiment enables us also to independently estimate the proper MEGNO 
integration time and to "calibrate" the scale of the diffusion rate,
$\sigma_0$. Comparing  the MEGNO and $\sigma_0$-maps, we conclude that for
$\log \sigma$ less than about  $(-7)$--$(-8)$ the motion can be considered as
regular (also compare the maps in the left  column of Fig.~\ref{fig:fig11}).
Let us note that both indicators are in an excellent accord and the diffusion
rate calculated by the MFT algorithm seems to be even more sensitive to an
unstable motion than the MEGNO is.

Using a similar numerical setup, we have investigated the region III. However, 
due to an extended range of the semi-major axes, we have divided it into two parts: the 
region IIIa between 0.5~AU and 0.97~AU and the region IIIb between 0.9~AU and 3.9~AU 
(a distance comparable to the radius of the Asteroid Belt in the Solar System). The maps 
of MEGNO and the diffusion rate, $\log \sigma$, in the $(a_0,e_0)$-plane for the region IIIa 
(obtained after the time $2T \simeq 32,000$~P$_{\idm{C}}$), are shown in the middle column of 
Fig.~\ref{fig:fig11} while the maps for the region IIIb ($2T \simeq 64,000$~P$_{\idm{C}}$) are 
shown in the right column. For the region IIIa, in both maps, we can clearly recognize the
collision line with the outermost planet while the crossing line with the planet B is in 
the zone of strong chaos. A number of MMRs appear as narrow vertical strips. This test shows 
that for a moderately low initial eccentricity, $e_0$, the stable zone is extended and begins 
just beyond the orbit of the planet C. Obviously, with a growing $e_0$ the border of the
stable zone shifts toward a larger $a_0$ (at the distances of about 1~AU the zone of 
stability reaches $e_0 \simeq 0.5$). For the region IIIb ($a_0 \in [0.9,3.9]$~AU), the 
results are shown in the right column of Fig.~\ref{fig:fig11}. For efficiency reasons, in 
this test the step in $e_0$ is $0.1$ (the resolution in $a_0$ was left relatively high, 
at $0.005$~AU). The collision lines are clearly present. Otherwise,
this zone is mostly regular even for very large $e_0>0.5$.  Similarly to the
region IIIa, the scans reveal some extremely narrow unstable MMRs, 
most of them with the two outer planets.

Finally, we examined the stability of orbits inclined to the mean orbital
plane of the system. In this experiment we tested the motion of particles in
the region III. The results for the region IIIa are shown in Fig.~\ref{fig:fig13}
(a number of additional scans, not shown here, allow us to extrapolate the results
obtained for this region to the region IIIb). In the first two experiments, we scanned 
the $(a_0,e_0)$ space for the initial $i_0$  set to $75^{\deg}$ and $87^{\deg}$, 
respectively. These inclinations, taken relative to the mean orbital plane of the \pstar 
system, are comparable to the inclinations of the Kuiper belt objects in the Solar System.
Again, the collision lines and the net of MMRs clearly appear in both the MEGNO and 
$\sigma_0$ scans which are shown in the the left and the middle panels of
Fig.~\ref{fig:fig13}.
The zone of strong chaos covers the crossing zones with the planets B and C while the 
collision line with the planet A is much more narrow and separated by a quasi-regular area 
in which $\sigma_0 \simeq 10^{-6}$. Such an effect has been already observed by 
\cite{Robutel2001} --- for higher inclination the close encounters with the planets are
less frequent than for moderately small inclinations which explains the smaller extent 
of the strong chaos. Moreover, after sufficiently long time the massless bodies will be 
removed from above the collision lines, excluding cases when the probes are trapped within 
stable MMRs. In the next test illustrated in the right panel of Fig.~\ref{fig:fig13},
$e_0$ was set to 0 and the initial inclination $i_0$ was varied in the wide 
range $[10^{\deg},90^{\deg}]$. This experiments allow us to generalize the results 
of the previous scans. The stability of inclined particles is basically independent on 
the initial inclination, at least for moderately small eccentricities.
  
One should be aware that the results of the above analysis should be treated as 
preliminary ones. Our study is restricted both to the short-term dynamics 
and a small part of the possible volume of the parameter space. However,  
the timescale of the integrations is long enough to detect the most unstable 
regions in the phase space and to point out the regions where the particles 
can be long-term stable. Also in some chaotic regions the particles can still 
persist over a very long time but this can be verified only by direct integrations. 
Let us also note that in order to derive rigorous estimates of the {\em
proper} mean motion frequencies one has to employ angle-action like
coordinates, e.g., the Poincar\'e coordinates \citep{Laskar1995} or the
Jacobi coordinates. Our integrations for mass-less particles were
carried out in the Keplerian, astrocentric coordinates and the osculating
elements analyzed by MFT are inferred from these non-canonical coordinates.
Due to very small masses involved, the effects caused by the use of
non-canonical coordinates are negligible.

\section{Conclusions}

In this work we carry out numerical studies of the stability of the \pstar
system using the initial condition determined by \cite{Konacki2003}. The 
long term integrations utilizing the symplectic integrators, extended over
1~Gyr, do not reveal any secular changes in the semi-major axes,
eccentricities and inclinations of the planets. Using the notion of the
Angular Momentum Deficit (AMD), we do not find any substantial exchange  of
the angular momentum between the innermost planet and the pair of the outer, 
bigger planets B and C. The AMD of the planet A is negligible when compared
to the  AMD of the B-C pair. This is very different from the case of the
inner Solar System in which the variations of the AMD of Mercury are the most
significant ones. The \pstar system has the MEGNO signature typical
for a strictly regular, quasi-periodic configuration. The two outer planets
are close to the 3:2 mean motion resonance and are orbitally tightly coupled.
The presence of the secular apsidal resonance is quite typical for such
system as demonstrated by \cite{Pauwels1983} and recently by \cite{Lee2003} and 
\cite{Michtchenko2004}. The semi-amplitude of the critical argument is about
$50^{\deg}$ and it persists in a wide range of the orbital initial
parameters. The SAR in the \pstar system is yet another such case among
extrasolar planetary systems \citep{Ji2003}. 

The neighborhood of the nominal initial condition is investigated by calculating
the MEGNO signature in a few representative regimes of the semi-major axes and 
eccentricities of the planets. These stability maps reveal that the
nominal initial condition is located in an extended stable zone, relatively
far from any strong instabilities of the motion. However, numerous weak mean 
motion resonances can be found in close proximity to the nominal positions of 
the planets. These are both 2-body resonances between the planets (like the 31:21~MMR
between B and C) and 3-body resonances, among which the 3:-9:14~MMR
seems to be the most relevant one. Their potential influence on the motion could
be investigated if the initial condition of the system was refined using
a more accurate model of the dynamics, possibly including relativistic 
effects. However, it would most likely require much more precise TOA
measurements than currently available.

These factors allows us to state that the \pstar system is orbitally stable 
over the Gyr time scale. In our experiments, there are no signs of a potential 
instability except for a very slow divergence of the MEGNO  in few of the 
tests. This divergence of MEGNO corresponds to the Lyapunov time of about  
$\simeq 1$~Gyr. We believe that it has only a numerical character.
We are aware that an alternative analytical study of the \pstar dynamics  is
possible (also  thanks to the accurate determination of the initial
condition). Such an approach has been proposed in \cite{Malhotra1989} and
\cite{Malhotra1993}. Our numerical  investigations can certainly be treated
as a complement to any future analytical studies of the system.

Using the MEGNO analysis, we found dynamical limits on the unconstrained
elements of the planet~A. Due to the destabilizing effect of the
Kozai resonance, the nodal line of this planet cannot be separated by more
than about $\pm60^{\circ}$ from the nodes of B and C. Otherwise, the
eccentricity of A would be excited to large values permitting close
encounters or collisions with the planet B. It constitutes a strong 
dynamical argument that the orbital plane of the planet~A indeed
coincides with the mean orbital plane of the system.

Finally, using the MEGNO and the MFT, we investigate the dynamics of massless
particles in the \pstar system in the framework of the restricted model. In
the numerical experiments, we find a stable zone between the planets A and B
extending for initially small eccentricities from 0.19~AU to 0.25~AU from the
pulsar. There are no stable areas between the planets B and C. Beyond the
orbit of the planet  C, the stable zone begins already outside of its orbit.
We find that the massless particles  can move on stable orbits under the
condition that their initial eccentricities and  semi-major axes are located
under the collision lines with the planets. The dynamics of  massless
particles is basically independent on the their initial inclinations. Beyond
1~AU,  the motion appears to be stable except for the areas of narrow MMRs
with the planets  B and C. It is an encouraging result supporting the search
for possible small bodies  contained in a dust or Kuiper belt type disk
around the \pstar.

\section{Acknowledgments}
We are indebted to the anonymous referee for many invaluable suggestions and
a detailed review which improved the manuscript.
We thank Philippe Robutel for a discussion and helpful remarks.
K.~G. is supported by the Polish Committee for Scientific Research, 
Grant No.~2P03D~001~22. M.~K. is a Michelson Postdoctoral Fellow.
\bibliographystyle{apj}
\bibliography{ms}
%


%
%

\figcaption[f1]{
The orbital evolution of the nominal \pstar system and its MEGNO signature.
Panel~(a) is for the semi-major axes. Panel~(b) is for the orbital
inclinations. Panel~(c) is for the eccentricities. Panel~(d) illustrates
the secular apsidal resonance between the planets B and C. Panels~(f) and~(e)
are for the time-evolution of MEGNO and its mean value, $\Ym(t)$ 
(a few representative evolutions for different choices of the initial 
tangent vector are shown). The orbital evolution is given in terms of
the heliocentric canonical elements related to the Poincar\'e coordinates.}

\figcaption[f2]{
The dynamics of the \pstar system in the 1~Gyr integration. Panel~(a) is
for the critical argument of the secular apsidal resonance during
1~Gyr. Panel~(b) is for the AMD. Panels~(c) and (d) are for
the argument $\theta_1=\varpi_{\idm{A}}-\varpi_{\idm{B}}$. Panel~(c) is
for the first 0.2~Myr while panel~(d) is for the end of the 1~Gyr
period. 
The solid line in these plots denotes the eccentricity of the innermost planet
multiplied by $10^4$. Panel~(e) shows the SAR between the planets B and
C in the space of ($\varpi_{\idm{A}}-\varpi_{\idm{B}},e_{\idm{C}}$).  The
same plot for $\theta_1$ and $e_{\idm{A}}$ is shown in panel~(f). 
The orbital evolution is given in terms of the heliocentric canonical elements
related to the Poincar\'e coordinates.}

\figcaption[f3]{
The dynamical environment of the \pstar planets in the space of the
semi-major axes. The plots are for the one-dimensional MEGNO scans along the 
semi-major axis of the planet A, B and C. All the other initial elements  are
fixed at their nominal values (see Table~\ref{tab:tab1}). The resolution  of
the scans is $5 \cdot 10^{-6}$~AU and $10^{-6}$~AU (the later for the
close-up scans for the planets A and C). Labels mark the positions of 
the MMRs between the planets: the upper plot and the second  plot for the A-B
pair the third, fourth and fifth panel from the top for the B-C
pair.  The upper plot is a magnification of the scan for the planet~A, the
bottom plot  is a magnification of the scan for the planet~C. Big dots mark
the nominal positions of the planets.
}

\figcaption[f4]{
The left column shows the evolution of the critical arguments of the MMRs close
to the planet C (see the text for explanation). The panels in the right column
are for MEGNO. The Lyapunov time about 3000~yr for the C31:B21~MMR is
estimated through a linear fit to the MEGNO plot. }

\figcaption[f5]{
The MEGNO scan along $a_{\idm{C}}$ for different masses of the host
star. The upper scan is for $m_{\star}=0.95 M_{\idm{psr}}$ and the lower for
$m_{\star}=1.05 M_{\idm{psr}}$ where $M_{\idm{psr}}$ is the canonical mass
of the pulsar, 1.4~$M_{\sun}$.
}

\figcaption[f6a,f6b,f6c]{
The MEGNO
stability maps for the configuration given in Table~\ref{tab:tab1}. The left
panel is for the $(a_{\idm{A}},e_{\idm{A}})$-plane.  The middle panel is for
the $(a_{\idm{B}},e_{\idm{B}})$-plane. The right panel is for the
$(a_{\idm{C}},e_{\idm{C}})$-plane. 
The position of the nominal system is marked 
by the two intersecting lines.
}

\figcaption[f7]{
The MEGNO
stability map for the configuration given in Table~\ref{tab:tab1}. 
The scan is for the $(e_{\idm{B}},e_{\idm{C}})$-plane.  
The position of the nominal system is marked 
by the two intersecting lines.
}

\figcaption[f8]{
The semi-amplitude $\theta^{\idm{max}}$ of the SAR in the 
$(a_{\idm{B}},e_{\idm{B}})$-plane. 
The position of the nominal system is marked by the two 
intersecting lines. 
}

\figcaption[f9]{
The stability map in the $(i_{\idm{A}},\Omega_{\idm{A}})$-plane
(the left panel) and the maximal eccentricity of the planet~A
attained during the integration time (the right panel).
The resolution of the plot is $1^{\circ}\times1^{\circ}$.
The position of the nominal system is marked by the two
intersecting lines. 
}

\figcaption[f10]{Kozai resonance for a configuration corresponding
to the initial $i_{\idm{A}}=50^{\circ}$ and $\Omega_{\idm{A}}=-75^{\circ}$.
Panel (a) shows the changes of the eccentricity and inclination 
of the planet A relative
to the invariant plane of the system. Panel $g_{\idm{a}}$ is for the 
argument of periastron, measured with respect to the invariant plane.
Panel for $Y(t)$ shows the temporal variations of MEGNO and its mean value,
$\Ym$.
}

\figcaption[f11ab,f11aa,f11bb,f11ba,f11cb,f11ca]{
The stability maps for the massless particles in different regions of the 
\pstar system. The left column is for the region~II (between the planets A nd B), 
the middle panel is for the region~IIIa (beyond the planet C, up to 1~AU)  and the
right panel is for the region~IIIb (up to 3.9~AU). The upper maps are for
MEGNO, the bottom maps are for the diffusion rate $\log \sigma_0$. The
resolution of the scans is $0.0008\mbox{AU}\times 0.005$,
$0.0025\mbox{AU}\times 0.005$ and $0.005\mbox{AU}\times 0.1$, respectively.
Collision lines with the planets are marked with white curves.
}

\figcaption[f12]{
The MEGNO scan along the semi-major axis $a_0$ of the massless particles
moving between the planet A and B (thick line) and their survival times over
15~Myr integrations (represented by thin vertical lines).
}

\figcaption[f13cb,f13ca,f13ab,f13aa,f13bb,f13ba]{
The stability maps for the massless particles moving in the region IIIa 
of the \pstar system for different initial inclinations. The left column is for the
the initial  inclination $i_0=75^{\deg}$, the middle column is for $i_0=87^{\deg}$. 
The right panels are for $e_0=0$ and $i_0 \in [10^{\deg},90^{\deg}]$. The upper 
maps are for MEGNO, the bottom maps are for the diffusion rate $\log \sigma_0$. 
The resolution of the scans is $0.0023\mbox{AU}\times 0.04$, 
$0.0019\mbox{AU}\times 0.01$,
and $0.0019\mbox{AU}\times 2^{\deg}$, respectively.
}


%
%
\setcounter{figure}{0}
%
%

\begin{figure*}[th]
\centering
\hbox{    
\includegraphics[]{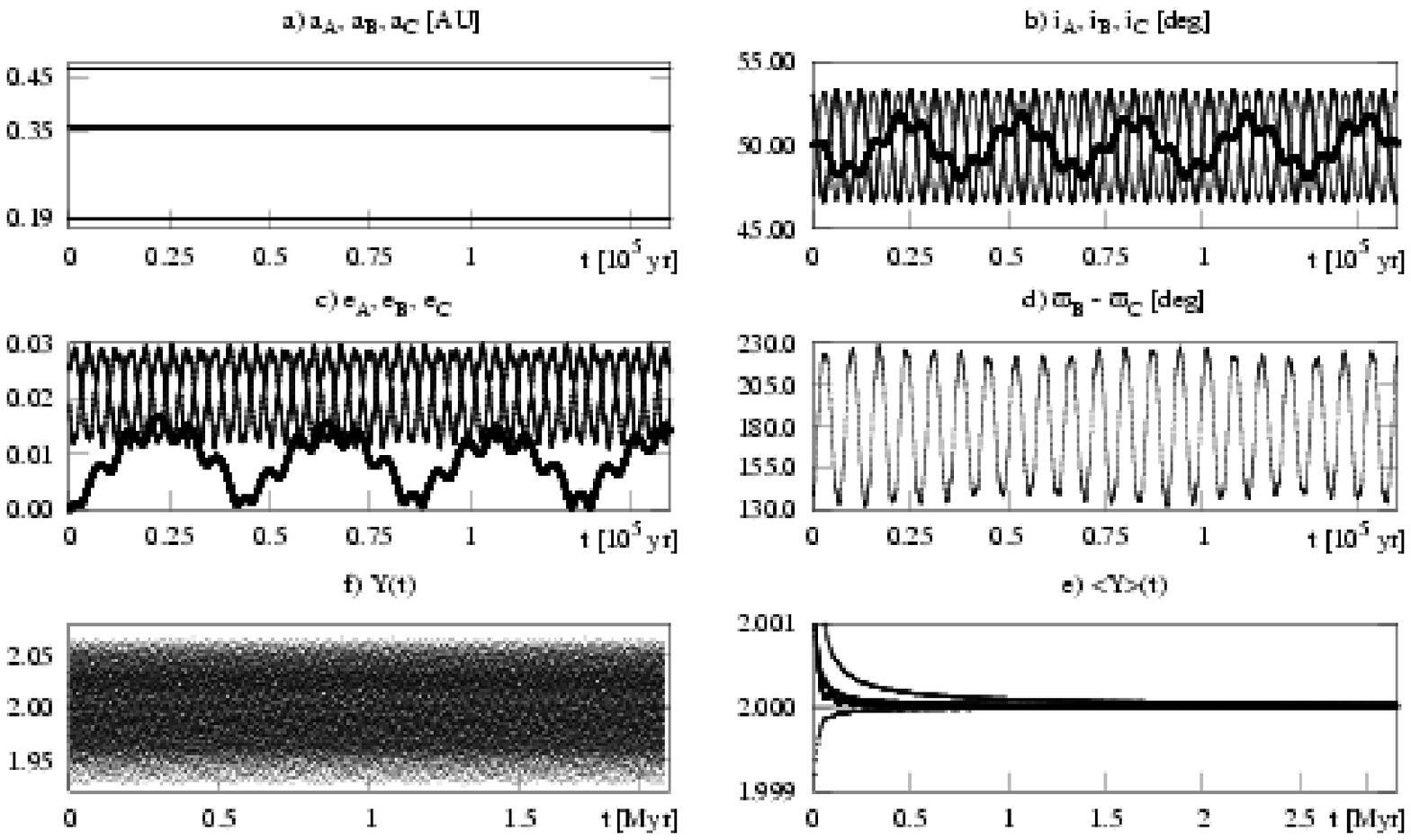}
     }
\caption{}
\label{fig:fig1}
\end{figure*}

%
%

\begin{figure*}[th]
\centering
\hbox{    
\includegraphics[]{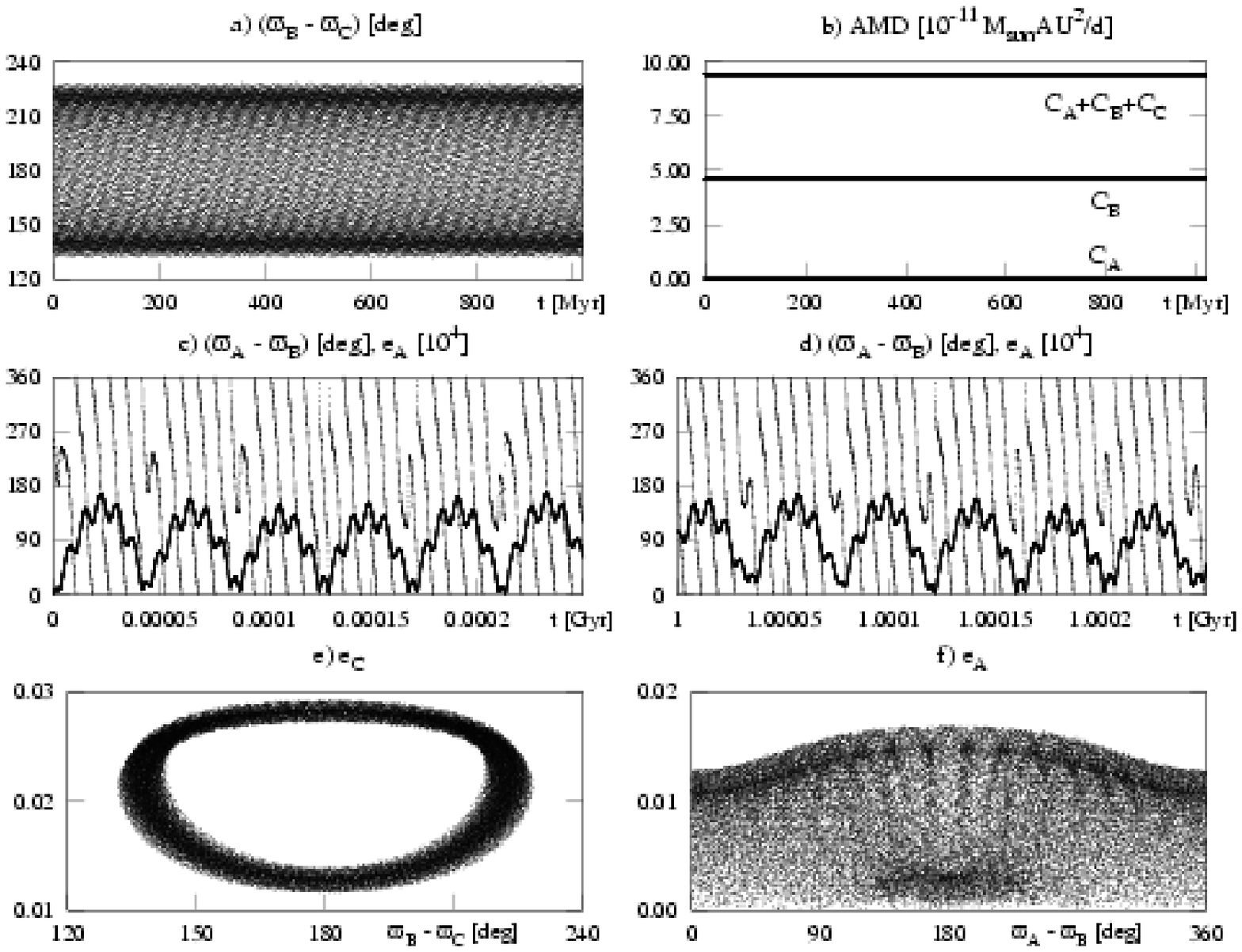}
     }
\caption{}
\label{fig:fig2}
\end{figure*}

%
%

\begin{figure*}[th]
\centering
\hbox{ 
\includegraphics[]{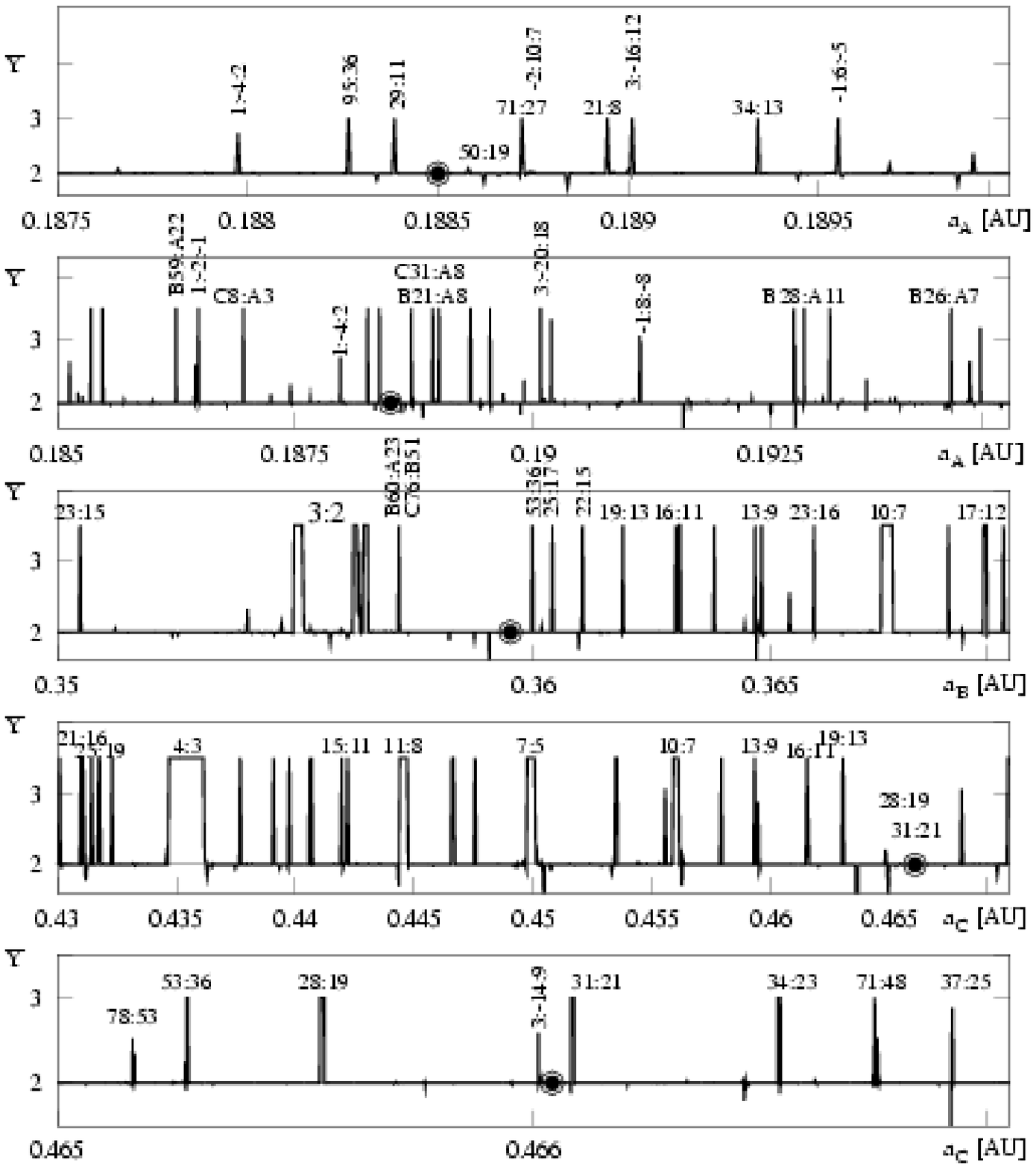}   
     }
\caption{}
\label{fig:fig3}
\end{figure*}

%
%

\begin{figure*}[th]
\centering
\hbox{\includegraphics[]{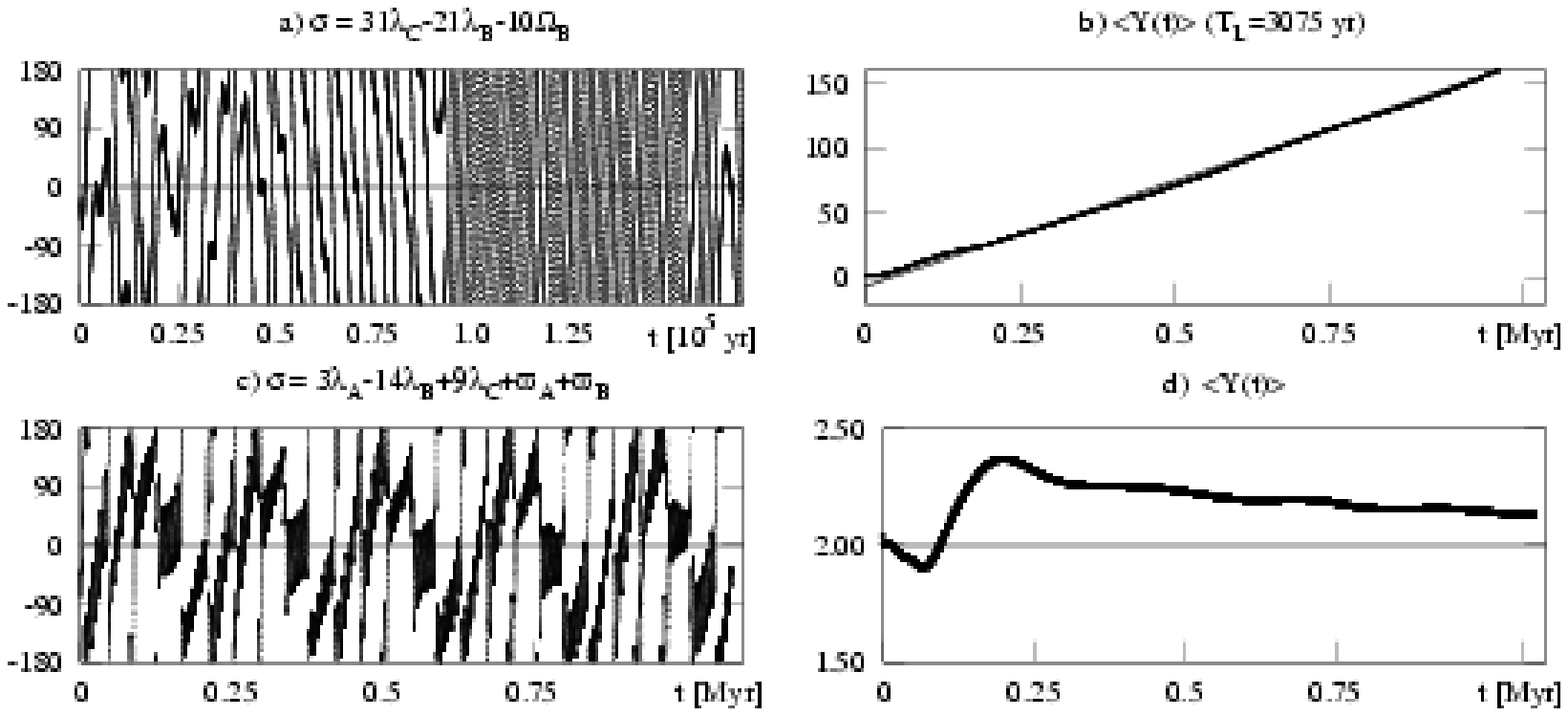}}
\caption{}
\label{fig:fig4}
\end{figure*}

%
%

\begin{figure}[th]
\centering
\hbox{\includegraphics[]{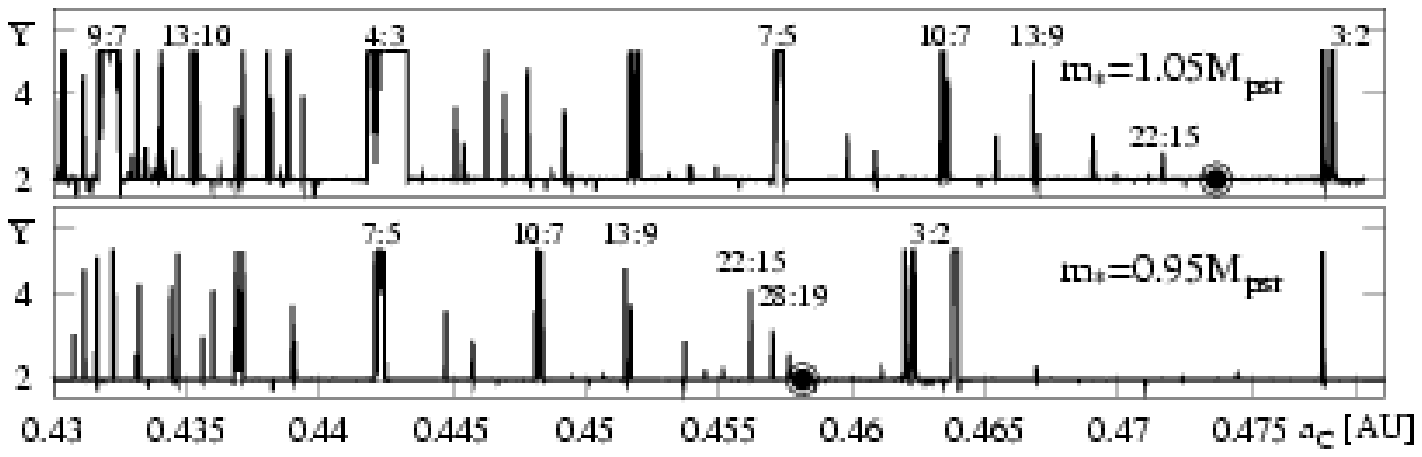}}
\caption{}
\label{fig:fig5}
\end{figure} 

%
%

\begin{figure*}[th]
\centering
\hbox{
\includegraphics[]{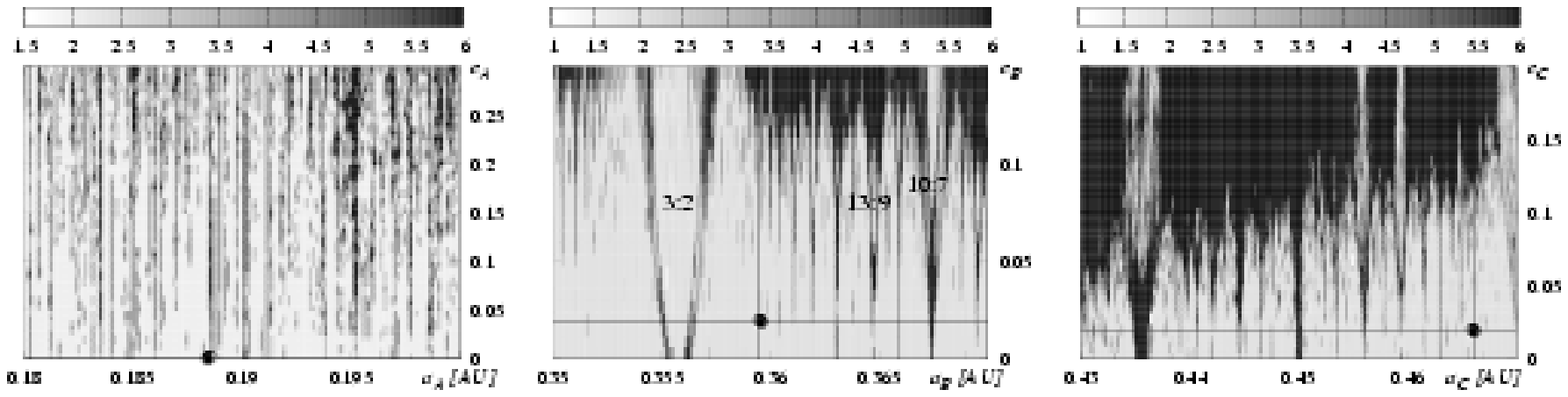} 
}
\caption{}
\label{fig:fig6}
\end{figure*}

%
%

\begin{figure}[th]
\centering
\hbox{\includegraphics[width=9cm]{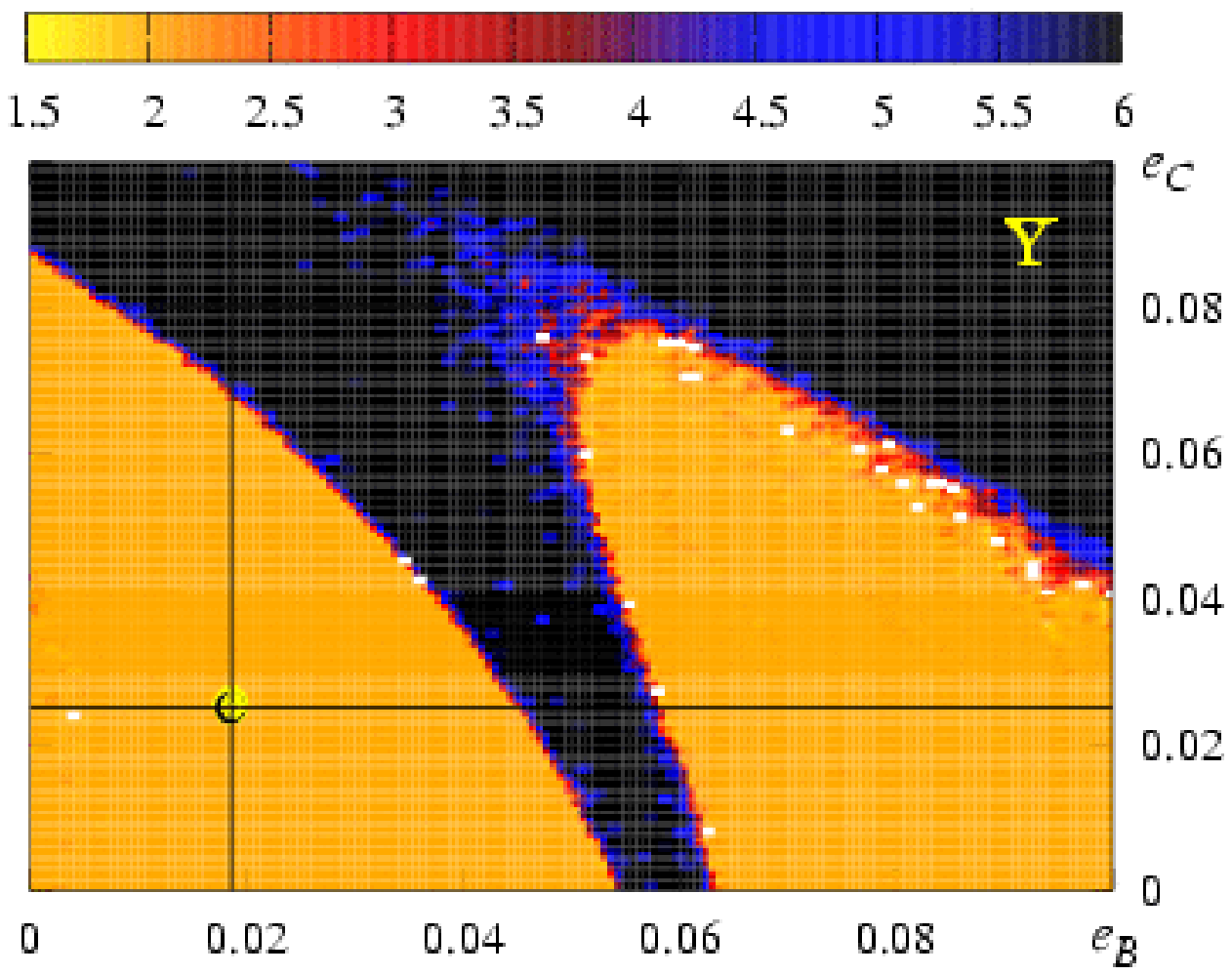}}
\caption{}
\label{fig:fig7}
\end{figure}

%
%

\begin{figure}[th]
\centering
\hbox{\includegraphics[width=9cm]{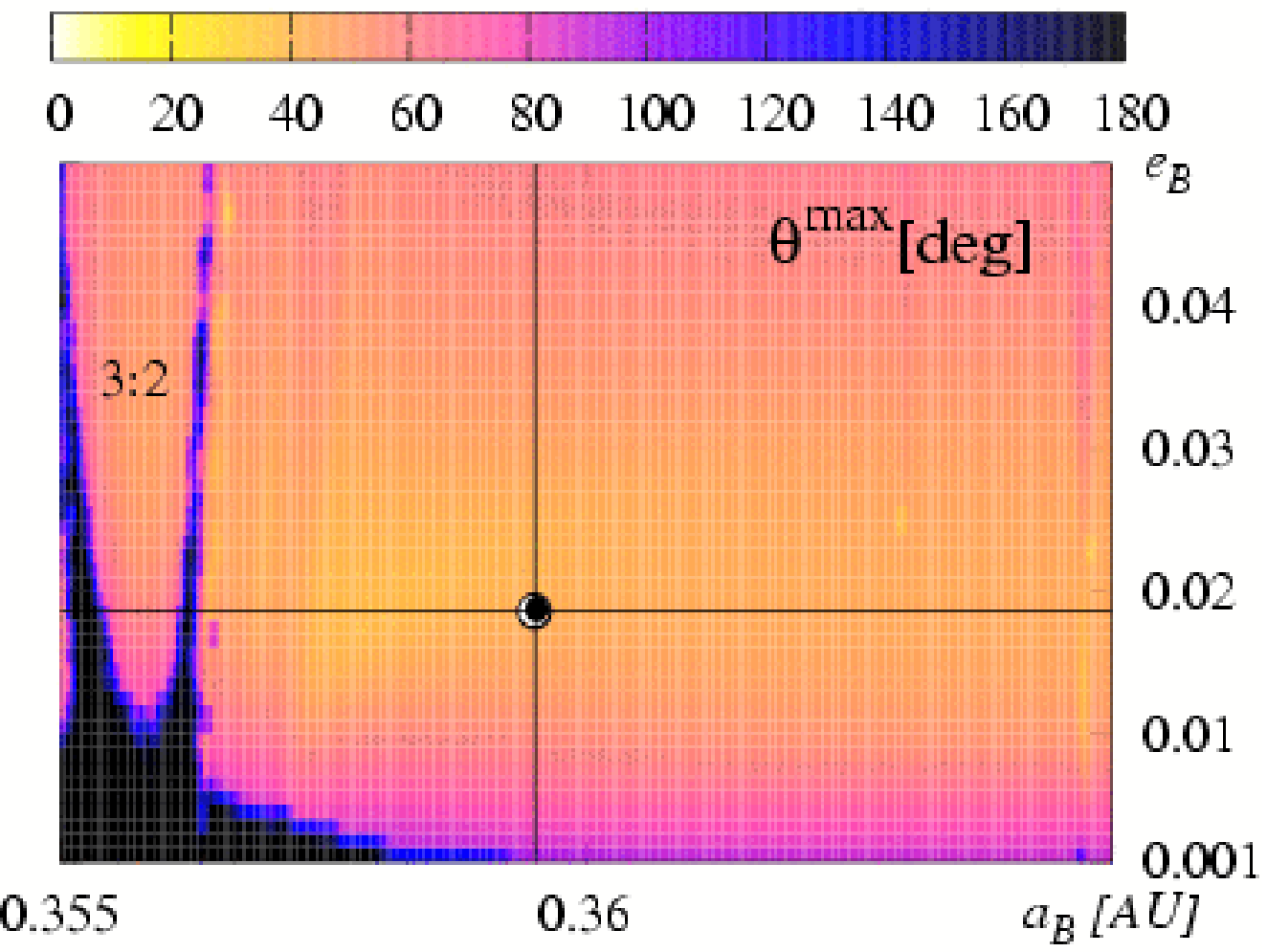}}
\caption{}
\label{fig:fig8}
\end{figure}

%
%

\begin{figure}[th]
\centering
\hbox{
\includegraphics[]{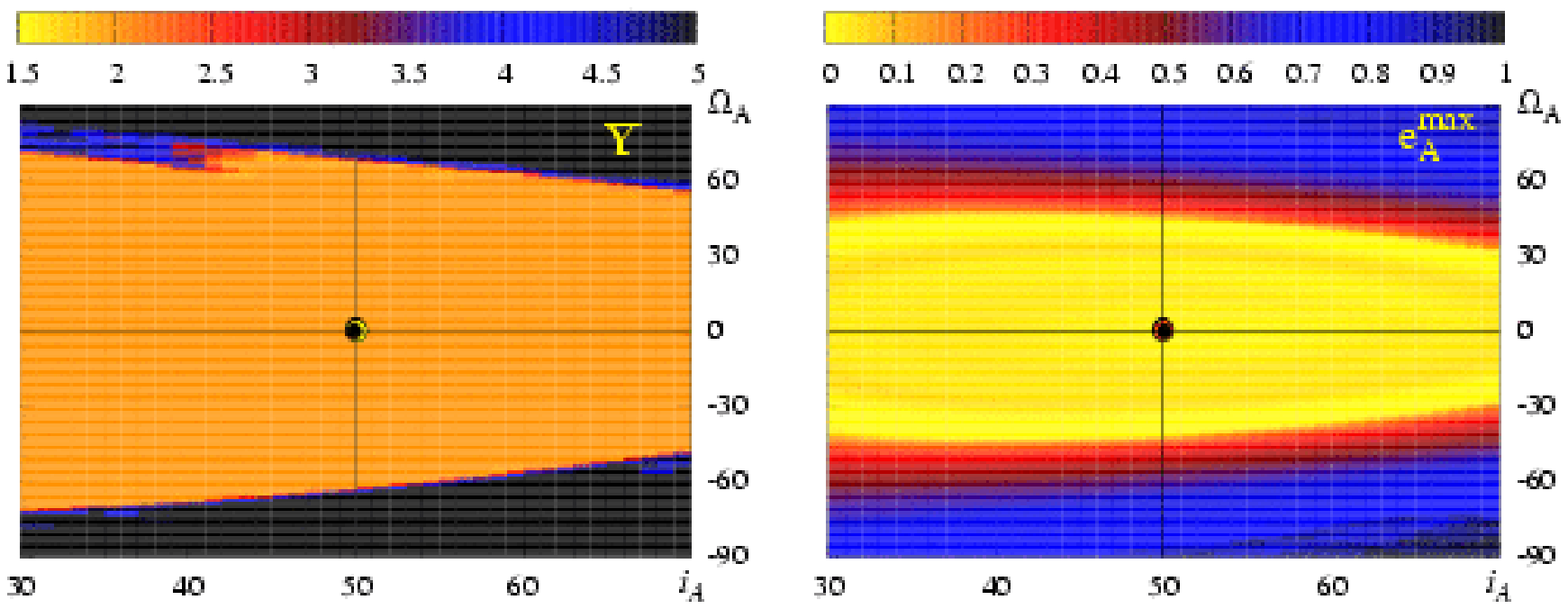}
}
\caption{}
\label{fig:fig9}
\end{figure}

%
%

\begin{figure*}[th]
\centering
\hbox{\includegraphics[]{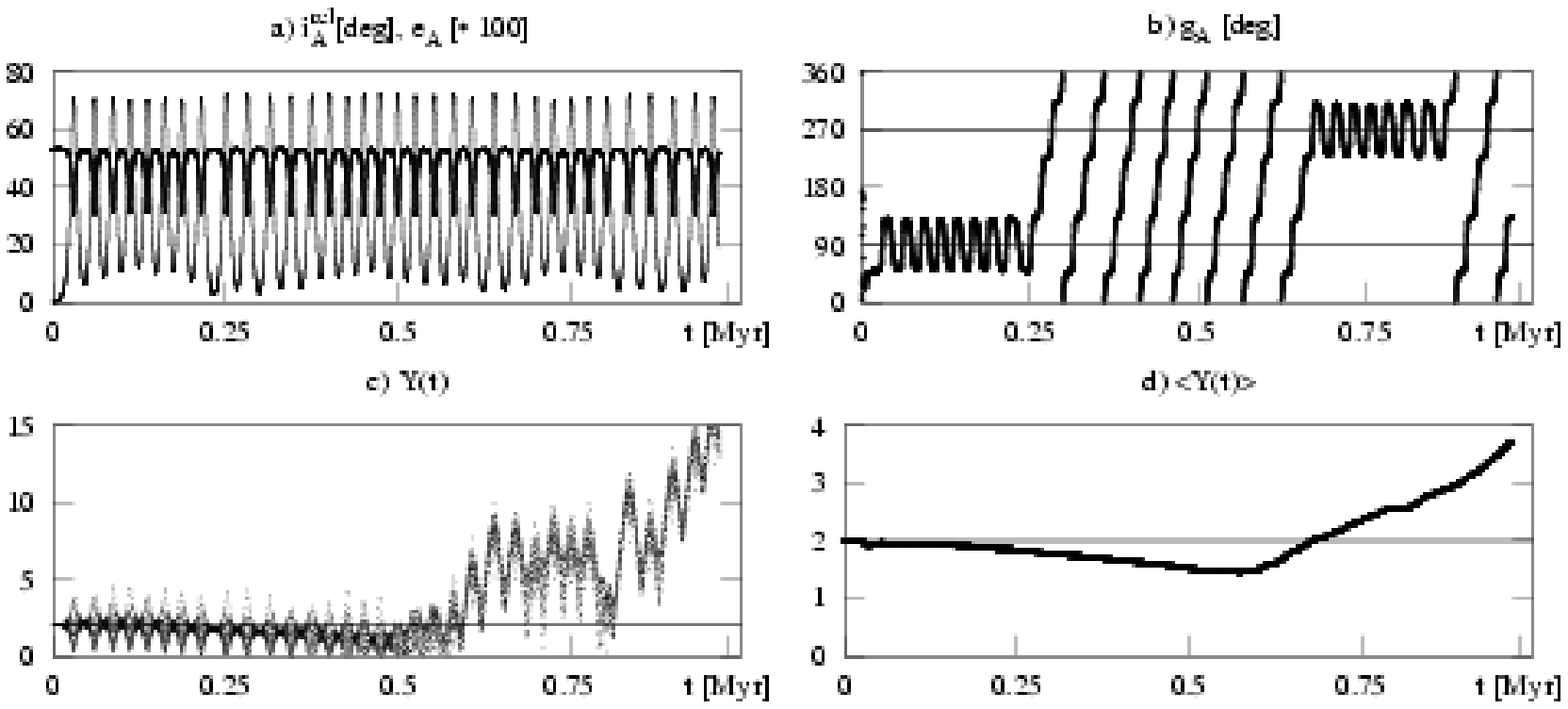}}
\caption{}
\label{fig:fig10}
\end{figure*}

%
%

\begin{figure*}[th]
\centering
\hbox{
\includegraphics[]{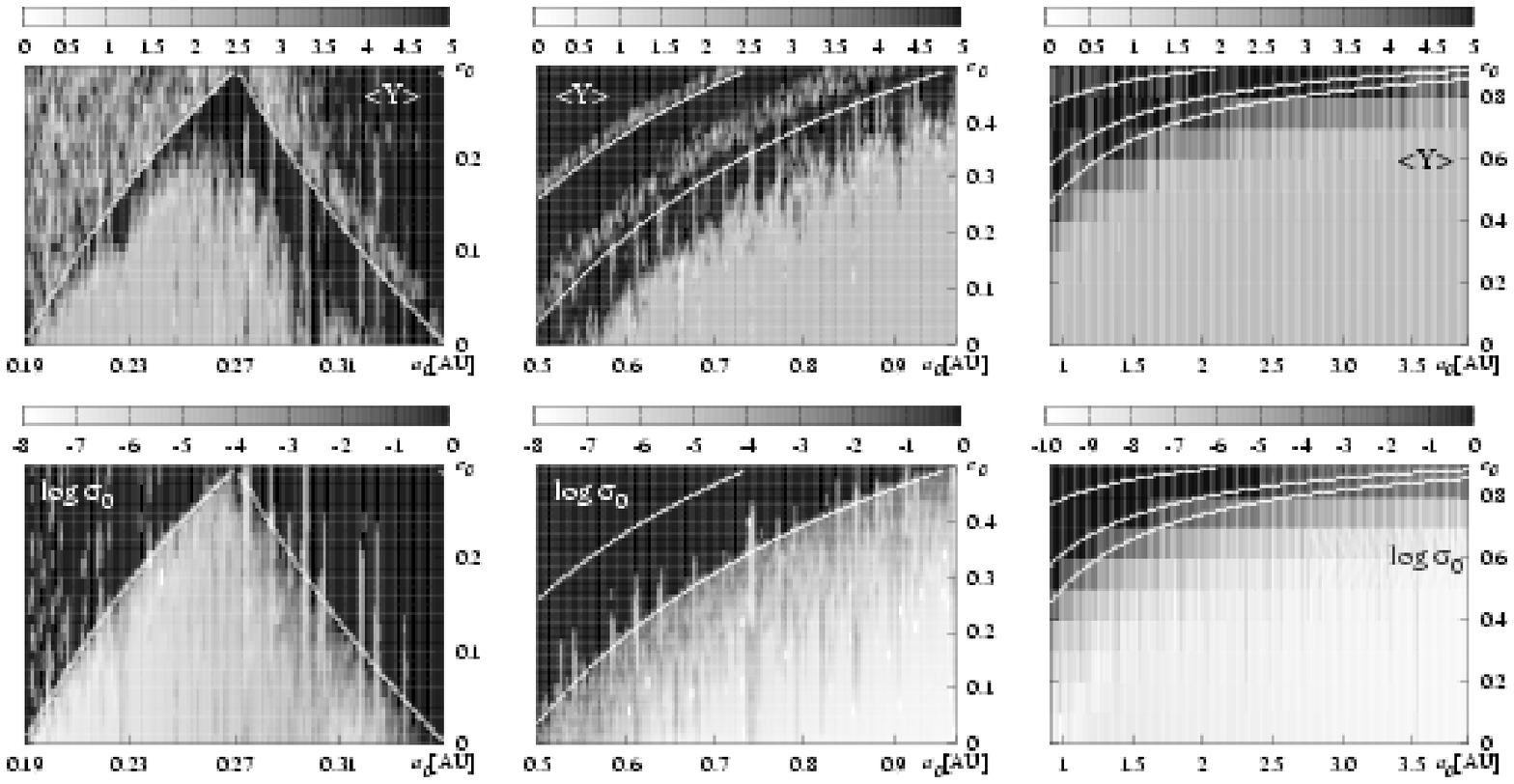}
}
\caption{}
\label{fig:fig11}
\end{figure*}

%
%

\begin{figure*}[th]
\centering
\hbox{\includegraphics[]{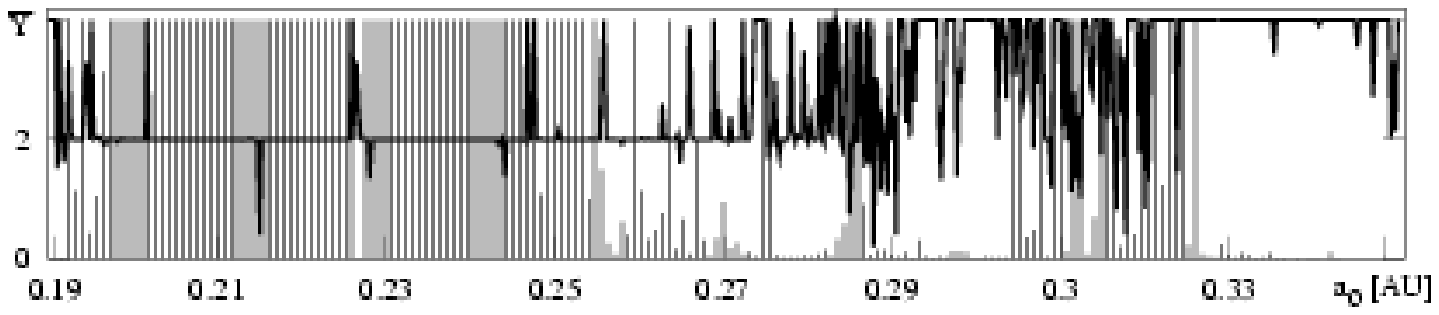}}
\caption{}
\label{fig:fig12}
\end{figure*}

%
%

\begin{figure*}[th]
\centering
\hbox{
\includegraphics{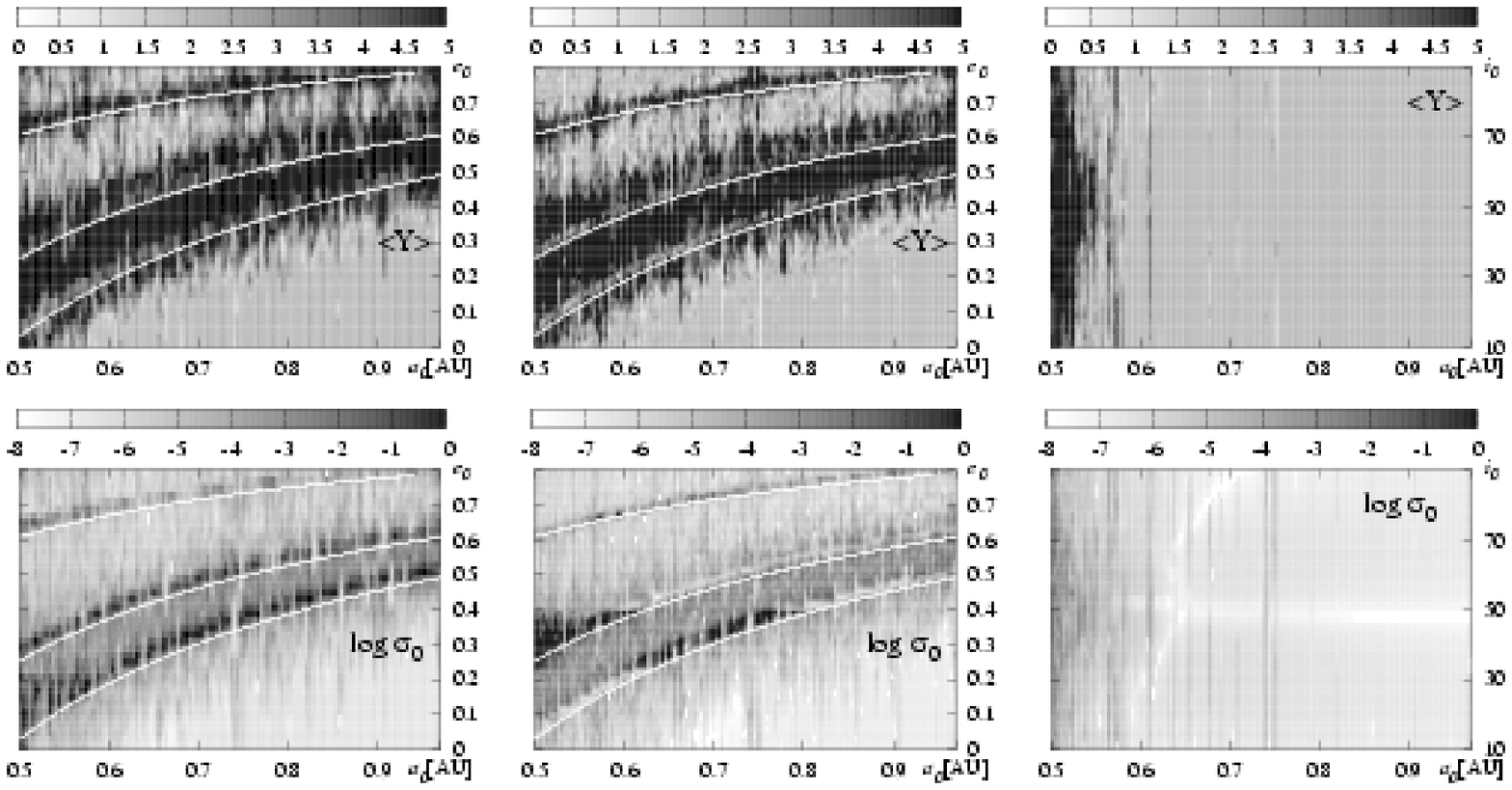}
}
\caption{}
\label{fig:fig13}
\end{figure*}


%
%

%
%

\begin{deluxetable}{lccccccc}
\tablewidth{400pt}
\tablecaption{
Astrocentric osculating orbital elements of the planets in \pstar planetary system at 
the epoch MJD=49766.50.\tablenotemark{a}}
\tablehead{\colhead{Planet} & \colhead{Mass [M$_{\oplus}$]} & \colhead{$a$~[AU]} & 
           \colhead{$e$}    & \colhead{$i$~[deg]}           & \colhead{$\Omega$ [deg]} & 
	   \colhead{$\omega$ [deg]} & \colhead{$M$ [deg]}}
\startdata
A &  0.019  & 0.18850 & 0.0000 & 50.00 & 0.00 &   0.0 & 14.25 \\
B &  4.250  & 0.35952 & 0.0186 & 53.00 & 0.00 & 250.4 &  5.41 \\
C &  3.873  & 0.46604 & 0.0252 & 47.00 & 3.26 & 108.3 &  3.66 \\
\enddata
\label{tab:tab1}
\tablenotetext{a}{
The semi-major axis, $a$, the eccentricity, $e$, 
the inclination, $i$, the longitude of ascending node, $\Omega$, 
the argument of periastron, $\omega$, and the mean anomaly, $M$, of the planets in \pstar planetary system at 
the epoch MJD=49766.50 are derived from the best-fit orbital parameters 
by \cite{Konacki2003}. The mass of the central star is equal to $1.4~\mbox{M}_{\sun}$.}
\end{deluxetable}

\begin{deluxetable}{cccrrrr}
\tablewidth{400pt}
\tablecaption{Parameters for the $(h,k)$ secular solution obtained by means 
of the Laplace-Lagrange theory.
Amplitudes $e_{\idm{p},i}$ are multiplied by $10^{3}$.}
\tablehead{ \colhead{Mode [$i$]} & 
\colhead{$g$ [arcsec/yr]} & \colhead{$\beta$~[deg]} &
\colhead{$e_{\idm{A},i}$}  & \colhead{$e_{\idm{B},i}$} & 
\colhead{$e_{\idm{C},i}$} & Period [yr] }
\startdata
1 &  209.991 & 272.28 &  -2.075 & 21.148 & -20.311 & 6171.7 \\
2 &   44.713 & 140.17 &  -8.146 &  0.010 &   0.013 & 28984.7 \\
3 &   14.061 & 333.01 &  -6.928 & -7.937 &  -7.960 & 92169.0   
\enddata
\label{tab:tab2}
\end{deluxetable}

\begin{deluxetable}{cccc}
\tablecaption{Fundamental frequencies and periods in the \pstar system.}
\tablehead{ \colhead{Frequency (period) [$i$]} 
        & \colhead{1} & \colhead{2} & \colhead{3} 
}
\startdata
$n_i$~[deg/d] 
             &  14.249~(25.264~d)   & 5.410~(66.544~d) &  3.665~(98.218~d) \\
$g_i$~[arcsec/yr] 
           &  197.696~(98502.1~yr)    & 43.881~(29534.0~yr)   & 13.157~(6555.5~yr) \\
$s_i$~[arcsec/yr] 
             &  -44.023~(29439.3~yr) &  -203.108~(6380.9~yr) & 0 (-)
\enddata
\label{tab:tab3}
\end{deluxetable}

\end{document}